\documentclass[fleqn,usenatbib]{mnras}
\usepackage{newtxtext,newtxmath}
\usepackage[utf8]{inputenc}
\usepackage{float}
\usepackage{graphicx}
\usepackage{amsmath}
\usepackage{datetime}
\usepackage[normalem]{ulem}
\usepackage{times}
\usepackage[export]{adjustbox}
\usepackage{soul}
\usepackage{caption}
\usepackage{dirtytalk}
\usepackage{gensymb}
\usepackage{hyperref}

\title[MOBSTER - VII. Properties of CM stars from photometry]{MOBSTER - VII. Using light curves to infer magnetic and rotational properties of stars with centrifugal magnetospheres}
\author{I. D. Berry, M. E. Shultz, S. P. Owocki, A. ud-Doula}

\author[I. Berry et al.]{I. D. Berry,$^{1}$\thanks{email: ianbass@udel.edu}
M. E. Shultz,$^1$
S. P. Owocki,$^{1,2}$ 
A. ud-Doula$^3$
}

\date{%
    $^1$\textit{Department of Physics \& Astronomy, University of Delaware, Newark, DE 19716, USA}\\%
    $^2$\textit{Bartol Research Institute, University of Delaware, Newark, DE 19716, USA}\\%
    $^3$\textit{Penn State Scranton, 120 Ridge View Dr., Dunmore, PA 18512, USA}\\[2ex]%
}

\begin{document}

\maketitle

\begin{abstract}
    Early-type B stars with strong magnetic fields and rapid rotation form centrifugal magnetospheres (CMs), as the relatively weak stellar wind becomes magnetically confined and centrifugally supported above the Kepler co-rotation radius. CM plasma is concentrated at and above the Kepler co-rotation radius at the intersection between the rotation and magnetic field axis. Stellar rotation can cause these clouds of material to intersect the viewer's line-of-sight, leading to photometric eclipses. However, for stars with strong ($\sim 10\,{\rm kG}$) magnetic fields and rapid rotation, CMs can become optically thick enough for emission to occur via electron scattering. Using high-precision space photometry from a sample of stars with strong H$\alpha$ emission, we apply simulated light curves from the Rigidly Rotating Magnetosphere model to directly infer magnetic and rotational properties of these stars. By comparing the values inferred from photometric modelling to those independently determined by spectropolarimetry, we find that magnetic obliquity angle $\beta$, viewer inclination $i$ and critical rotation fraction $W$ can be approximately recovered for 3 of the 4 stars studied here. However, there are large discrepancies between the optical depth at the Kepler radius $\tau_{\rm K}$ expected from magnetometry, and the values required to match the observations. We show that $\tau_{\rm K}$ of order unity is needed to reasonably match the light curve morphology of our sample stars.
\end{abstract}


\section{Introduction} \label{Intro}
Hot, luminous and massive O and B stars have dense line-driven stellar winds \citep{1975ApJ...195..157C,1995ApJ...454..410G,vink2001}. A small but significant subset ($\sim$ 10$\%$) host surface magnetic fields \citep{2017MNRAS.465.2432G,2019MNRAS.483.3127S} with strengths on the order of hundreds of G to tens of kG \citep{2019MNRAS.487.4695S,2019MNRAS.489.5669P,2019MNRAS.483.3127S,Shultz_2019}. The vast majority of these magnetic fields show simple topologies, with most being well approximated by dipoles tilted with respect to the rotational axis \citep{2019A&A...621A..47K}. The combination of a dense ionized stellar wind and a strong magnetic field allows plasma to become magnetically trapped within field loops, forming a circumstellar magnetosphere \citep{1997A&A...323..121B,2002ApJ...576..413U}, which is forced into co-rotation with the magnetic field. As demonstrated strikingly in the prototypical magnetospheric hot star $\sigma$ Ori E, this leads to magnetic, emission line, and photometric variability locked to the rotational period of the star \citep{1978ApJ...224L...5L}.

The structure of such circumstellar magnetospheres depends strongly on the stellar rotation. Magnetic O stars tend to be slow rotators due to magnetic torque applied onto their very strong winds leading to rapid loss of angular momentum \citep{2009MNRAS.392.1022U,2013MNRAS.429..398P}. Slow rotation  means that the Kepler co-rotation radius ($R_{\rm K})$ (the point at which gravitational and centrifugal forces balance in plasma co-rotating with the stellar magnetic field) is well beyond the Alfv\'en radius ($R_{\rm A}$) (the maximum distance of closed magnetic field loops). As such, trapped stellar wind material lacks centrifugal support to counteract gravity, and so falls back down to the stellar surface on dynamical timescales. This is known as a \say{dynamical magnetosphere} \citep[DM;][]{2002ApJ...576..413U,2013MNRAS.429..398P}.

B-type stars are less luminous than O stars. A consequence of this is that their line-driven winds are not nearly as powerful and dense as those of O stars. This means that their Alfv\'en radii are typically much larger than those of O-type stars. The magnetic torque applied to the winds of B stars is also much less than that of O stars, leading to shorter periods of rapid rotation\footnote[1]{N.B. there is one O star with a centrifugal magnetosphere (CM), Plaskett's Star, which may have been spun up by binary mass transfer; \citealt{Grunhut_2012,2022MNRAS.512.1944G}. There are also several B stars without CMs, e.g. the ultra-slow rotator $\xi^1$ CMa; \citealt{2017MNRAS.471.2286S,2018MNRAS.478L..39S,2021MNRAS.506.2296E}}. B stars can rotate so rapidly that the Kepler co-rotation radius falls below the Alfv\'en Radius. In this scenario, material fed into the magnetosphere at or above $R_{\rm K}$ now has the centrifugal support to remain suspended above the surface, but is stopped from flowing further by closed magnetic field loops. This trapped plasma accumulates into dense clouds in the region between $R_{\rm K}$ and $R_{\rm A}$ forming a structure called a \say{centrifugal magnetosphere} \citep[CM;][]{2005MNRAS.357..251T,2008MNRAS.385...97U,2013MNRAS.429..398P}. Many B-type stars have $R_{\rm K}$ within half a stellar radius of the photosphere, rendering their magnetospheres almost entirely centrifugal. Rapidly rotating B stars are however typically young, as they are also spun down due to angular momentum loss from magnetic torque applied on their winds \citep[as shown by a comparison between observations and evolutionary models self-consistently accounting for magnetic braking;][]{2019MNRAS.485.1508S,Shultz_2019,2019MNRAS.485.5843K,2020MNRAS.493..518K}.

For B stars, the ratio between the energy densities of the magnetic field and the stellar wind outflow can be large enough such that the magnetic field is completely dominant over stellar wind material out to tens of stellar radii. This justifies the \say{rigidly rotating magnetosphere} (RRM) model developed by  \cite{2005MNRAS.357..251T} which finds the accumulation surface of a CM with a tilted dipole under the key assumption that the magnetic field remains completely rigid and unperturbed by the presence of circumstellar material out to large distances.

Photometric variation can be dominated by the presence of a CM, with the prototypical example of this found from the B2Vp star $\sigma$ Orionis E \citep[HD 37479, henceforth $\sigma$ Ori E;][]{1978ApJ...224L...5L}. This star shows photometric variability due to a cloud forced into co-rotation by $\sigma$ Ori E's $\sim$10 kG \citep{2012MNRAS.419..959O} magnetic field. Two major dips in brightness are visible during this star's 1.19 day rotation period, with photometric minima occurring simultaneously with phases of magnetic nulls (i.e. phases corresponding to a line-of-sight perpendicular to the magnetic field axis). \cite{1978ApJ...224L...5L} were the first to suggest that such variation is due to co-rotating, magnetically trapped plasma, with photometric minima occurring when material eclipses the star. This idea received strong support when \cite{2005ApJ...630L..81T} fit the light curve of $\sigma$ Ori E obtained by \cite{1977ApJ...216L..31H} with the RRM model. Further analysis of the light curve of $\sigma$ Ori E was done by \cite{2015MNRAS.451.2015O}, who employed an arbitrary rigidly rotating magnetosphere (aRRM) model based on a Zeeman Doppler Imaging map to simulate a light curve with contributions from surface spots. \cite{2015MNRAS.451.2015O} found that this model could match eclipse depth and width with a viewer inclination with respect to the rotation axis of $85^\circ$, but could not match out-of-eclipse variability even after including chemical spot contributions inferred from Doppler Imaging maps.

While RRM assumes a completely rigid magnetic field, regardless of the amount of material within the magnetosphere, in reality all magnetic fields have a finite strength, and hence a finite rigidity. This means that at some point, the centrifugal force acting on the accumulating plasma will overpower magnetic tension. This is known as centrifugal breakout (CBO), first suggested to be the mechanism by which plasma is ejected from stellar magnetospheres by \cite{1984A&A...138..421H}, who examined and rejected ambipolar diffusion as a plausible alternative. Further analysis is given in the Appendix of \cite{2005MNRAS.357..251T}, who provided scalings for density, timescale and breakout-limited asymptotic mass. 

\cite{Townsend_2013} challenged large-scale magnetosphere emptying via CBO events using the \textit{MOST} \citep[Microvariability and Oscillations in Stars;][]{2003PASP..115.1023W} satellite to measure photometric variability of $\sigma$ Ori E over a continuous observing window of three weeks. These observations found no evidence of large-scale magnetospheric emptying and reorganization as suggested by the magnetohydrodynamic (MHD) simulations conducted by \cite{2006ApJ...640L.191U,2008MNRAS.385...97U}. This motivated the development of a new model by \cite{2018MNRAS.474.3090O}, in which CM plasma transport is governed via a gradual \say{leakage} by diffusion across magnetic field loops. However, analysis of the full population of H$\alpha$-bright CM stars by \cite{2020MNRAS.499.5379S} and \cite{2020MNRAS.499.5366O} showed that the onset of emission, the emission strength scaling, and the H$\alpha$ emission line profile morphologies of in CMs of B-type stars can only be explained by CBO events. 

Magnetic B-stars often exhibit radio emission from  gyrosynchrotron and electron cyclotron maser emission. Recent radio observations by \cite{2021ApJ...921....9D} report the first observation of a \say{giant pulse} from the star CU Vir, larger than any seen previously, which they speculated could be linked to a large-scale CBO event since the enhanced emission came from both magnetic hemispheres. \cite{2021MNRAS.507.1979L} conclusively demonstrated that the previous wind-powered current sheet model for electron acceleration \citep[e.g.][]{2004A&A...418..593T} was inconsistent with the radio luminosities and mass-loss rates of magnetic stars, and found a rotationally-dependent luminosity scaling for gyrosynchrotron emission, suggesting CBO as a possible mechanism driving electron acceleration. This relationship was confirmed by the larger sample analyzed by \cite{2022MNRAS.513.1429S}, who also found a close correlation between H$\alpha$ emission strength and radio luminosity, further supporting CBO as a common mechanism.  \cite{2022MNRAS.513.1449O} then demonstrated that the gyrosynchrotron scaling relationship can be explained as a consequence of magnetic reconnection during CBO events, which accelerates electrons to the required semi-relativistic energies.

In their original analysis of CBO, \cite{2005MNRAS.357..251T} assumed that the magnetosphere was last emptied out at some arbitrary time in the past, with the density at some later time set by the wind feeding rate, which is proportional to the magnetic field strength. However, as demonstrated by \cite{Townsend_2013}, there is no evidence of the large-scale, magnetosphere-emptying breakout events in photometric time series, a conclusion supported by \cite{2020MNRAS.499.5379S}, who found H$\alpha$ profiles to be unchanged across more than 20 years. Since magnetospheric diagnostics are consistent with the visible material being always just at the breakout density, \cite{2020MNRAS.499.5379S} and \cite{2020MNRAS.499.5366O} speculated that CBO must be continuously occurring on small spatial scales. Calibrating their analysis with the MHD simulations performed by \cite{2006ApJ...640L.191U,2008MNRAS.385...97U}, \cite{2020MNRAS.499.5366O} found that CBO implies a radial surface density that varies with the square of the magnetic field strength. This leads to a steeper radial decline in density, but an overall higher density within the CM, than is obtained with the filling-time prescription adopted by \cite{2005MNRAS.357..251T}. This higher density means that electron scattering can lead to a significant and detectable amount of magnetospheric emission \citep{Berry_2022}.

Models of CMs have been used to explore photometric variation for a range of obliquities, inclinations and rotation periods. \cite{2008MNRAS.389..559T} demonstrated that prominent circumstellar eclipses can only be detected with highly tilted magnetic fields at high inclination with respect to the rotational axis. 

Magnetospheric eclipsing is not the only, or even the primary, mechanism by which rotationally modulated variation can occur with magnetic B-type stars. These stars are almost invariably chemically peculiar \citep[CP; e.g.][]{2006A&A...450..763K,2019MNRAS.483.2300S}, with $\sigma$ Ori E, a helium-rich star, falling into this category. Chemical peculiarities in such stars are not homogeneous across the surface, but instead are typically clumped together in abundance patches. These chemical abundance spots lead to periodic photometric variation as the star rotates \citep[e.g.][]{2001A&A...378..113R,2007A&A...470.1089K,2009A&A...499..567K,2013A&A...556A..18K,2019MNRAS.487.4695S}.

\cite{Berry_2022} conducted a similar study to \cite{2008MNRAS.389..559T}, covering a parameter space consisting of magnetic obliquity $\beta$, inclination with respect to the rotation axis $i$, critical rotation fraction $W$ and optical depth at the Kepler co-rotation radius $\tau_{\rm K}$. However, \cite{Berry_2022} also took into account emission from CMs via electron scattering, where \cite{2008MNRAS.389..559T} had only considered absorption. \cite{Berry_2022} used a  CBO modified \citep{2020MNRAS.499.5366O} and MHD calibrated \citep{2021mobs.confE..33U} density scaling for the RRM model, and found similar results to \cite{2008MNRAS.389..559T}, with the addition that CMs can become sufficiently optically thick (i.e. $\tau_{\rm K} \geq 1$) in the continuum to give up to 5$\%$ extra emission above the stellar flux due to electron scattering.


Despite magnetic B-type stars being far more numerous than magnetic O-type stars, to date magnetospheric analyses of photometric eclipsing have been performed for more O-type than B-type stars. Photometric variations due to scattering by the varying column density of DMs around magnetic O-type stars was explored by \cite{2020MNRAS.492.1199M}, who used the Analytical Dynamical Magnetosphere model developed by \cite{2016MNRAS.462.3830O} to model light curves obtained from the Optical Gravitational Lensing Experiment \citep[OGLE;][]{2015AcA....65....1U} of several Of?p stars in the Magellanic Clouds \citep{2015A&A...577A.107N}. While Galactic Of?p stars are invariably magnetic \citep[e.g.][]{2011MNRAS.416.3160W}, spectropolarimetric observations of their counterparts in the Magellanic Clouds were unable to detect their magnetic fields, and nor are plausible future observations likely to do so \citep{2020A&A...635A.163B}. Since the magnetic fields of such distant stars are beyond the range of contemporary spectropolarimeters, \cite{2020MNRAS.492.1199M} used the ADM model to infer the geometrical and magnetic properties of the extragalactic Of?p stars directly from their light curves. 

By contrast, RRM models have so far been applied to the light curves of only two magnetic B-type stars: $\sigma$ Ori E and Landstreet's Star (HD\,37776). The latter was analyzed by \cite{2022A&A...659A..37K}, who demonstrated using a multipolar RRM model that the complex, highly structured light curve of Landstreet's Star \citep[known to have a highly complex magnetic geometry and corresponding complex H$\alpha$ emission profile variability, e.g.][]{2011ApJ...726...24K,2020MNRAS.499.5379S} could be qualitatively reproduced with the superposition of surface chemical spots and magnetospheric clouds formed in a magnetosphere dominated by higher-order multipolar components in the region near the Kepler radius. 



Previous light curve models of CMs have been performed using forward modeling, with the light curve derived from a known or assumed magnetic field geometry \citep[e.g.][]{2005ApJ...630L..81T,2015MNRAS.451.2015O,2022A&A...659A..37K}. In this paper, we perform a blind test of the models, using light curves of four stars to derive geometrical, rotational and opacity parameters without reference to the values determined from spectroscopy, magnetometry and evolutionary models. These are then compared to the parameters determined by \cite{Shultz_2019} on the basis of spectropolarimetric measurement of the surface magnetic field, spectroscopic evaluation of atmospheric parameters, and fundamental stellar parameters inferred from evolutionary models. The targets -- $\sigma$ Ori E, HD 142184, HD 182180, and HD 345439 -- were selected on the basis that they have well-constrained magnetic, stellar, and rotational properties, and have available space-based photometry with light curves that appear to be magnetospherically dominated i.e.\ contributions from surface chemical spots can be plausibly neglected. Each light curve presents apparent eclipses that decrease the flux by at least $\sim 2\%$. Furthermore, three out of the four light curves show two apparent eclipses, including that of $\sigma$ Ori E. All four have extremely strong H$\alpha$ emission, amongst the strongest of the large sample of H$\alpha$-bright CM stars examined by \cite{2020MNRAS.499.5379S}. Each star shows variability in H$\alpha$ equivalent width \citep{2012MNRAS.419..959O,2012MNRAS.419.1610G,2013Rivinius,Wisniewski_2015}, which correlates closely with variability in the light curve, and traces to the presence of a CM \citep{2012MNRAS.419.1610G}. This extremely strong H$\alpha$ emission, and the finding by \cite{2015MNRAS.451.2015O} that the contribution of chemical spots to $\sigma$\,Ori\,E's light curve is negligible as compared to the magnetospheric contribution, underlie our assumption that the light curves of these stars can be reproduced with purely magnetospheric models. The goal of this study is to determine to what extent photometric analysis using the simplest possible RRM model (a tilted dipole) can recover the known magnetic, geometrical, and rotational properties. 

The work presented here uses photometry from three high-precision space photometers, one of which is the Transiting Exoplanet Survey Satellite \citep[{\it TESS};][]{2015JATIS...1a4003R}. These data are fit with our RRM-CBO models, which allow us to infer rotational properties solely from photometric time-series. This is a primary goal of the MOBSTER collaboration \citep[Magnetic OB(A) Stars with \textit{TESS}: probing their Evolutionary and Rotational properties;][]{2019MNRAS.487..304D}, where \textit{TESS} is used to increase the number of known rotational periods for magnetic chemically peculiar stars, which can then be utilized to infer evolutionary and magnetospheric properties of these stars.


Target selection and observations are described in Section \ref{obs}. Techniques for our model fitting are given in Section \ref{model}. Results for individual stars are provided in Section \ref{Results}. Implications of this work are discussed in Section \ref{Discussting}. Finally, a summary and conclusions are given in Section \ref{Summary}. 

\section{Target Selection and Observations} \label{obs}

\begin{figure}
    \centering
    \includegraphics[width=\columnwidth]{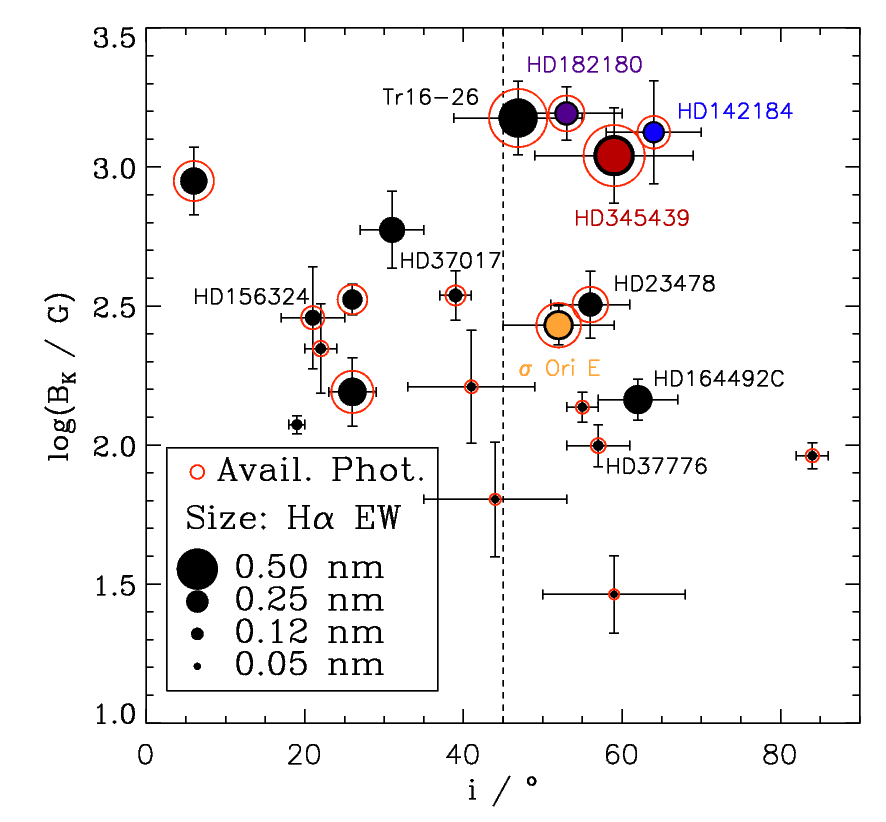}
    \caption{$\log{B_{\rm K}}$ as a function of the inclination angle $i$ for stars with CM-type H$\alpha$ emission. As indicated in the legend, symbol size is proportional to H$\alpha$ emission EW, and outlined symbols indicate stars with available space photometry. Stars studied in this work are highlighted and labeled. Other labeled stars are discussed in the text. The vertical dashed line indicates the approximate lower limit in inclination for eclipsing.}
    \label{fig:targets}
\end{figure}

In order to illustrate the relatively comprehensive nature of this study, Figure\ \ref{fig:targets} shows the logarithmic strength of the equatorial magnetic field at $R_{\rm K}$, $B_{\rm K} = B_{\rm eq}(R_\ast/R_{\rm K})^3$ (where $B_{\rm eq}$ is the surface equatorial magnetic field strength), as a function of the rotational axis inclination angle $i$, for stars with H$\alpha$ emission characteristic of line formation in a centrifugal magnetosphere. The majority of the stars were taken from the sample examined by \cite{2020MNRAS.499.5379S}, with three subsequently discovered additions (HD\,156424B, \citealt{2021MNRAS.504.4850S}, W\,601B, \citealt{2021MNRAS.504.3203S}, and Tr~16-26 \citealt{2022MNRAS.516.2812C}). Symbol size is proportional to H$\alpha$ emission equivalent width (EW), i.e.\ the difference between the observed and synthetic photospheric H$\alpha$ profile at the rotational phase of maximum emission strength; this quantity serves as a proxy for the size of the magnetosphere, and as discussed by \cite{2020MNRAS.499.5379S} has been corrected for dilution in the case of spectroscopic binaries. 

The vertical dashed line at $i = 45^\circ$ indicates the approximate lower limit for stars to be eclipsed by their CMs. Of those stars with large magnetospheres as inferred from their H$\alpha$ emission, only 7 are above this limit. For one, HD\,164492C, space photometry is not available \citep[and is likely to be difficult to interpret if it does become available, as this is a dim star in a very crowded field;][]{2017MNRAS.465.2517W}. Another, HD\,23478, has strong H$\alpha$ emission and an available {\em TESS} light curve, which has been examined by \cite{2022ApJ...924L..10J} and may show subtle signs of magnetospheric influence. However, its H$\alpha$ emission shows very low levels of variability \citep{2015MNRAS.451.1928S,2020MNRAS.499.5379S}, which is consistent with the very small $\beta$ angle (about $6^\circ$) inferred from its magnetic field measurements \citep{2015MNRAS.451.1928S,Shultz_2019}. Since little variation is expected in the light curve as a result of its magnetosphere, it is possible that much of the variability originates with photospheric spots, and we therefore chose to leave it out of the analysis. Tr~16-26 exhibits strong double-humped H$\alpha$ consistent, possesses a strong surface magnetic field of at least 11 kG, and displays two prominent eclipses in its {\em TESS} light curve which correspond to nulls in the longitudinal magnetic field curve, all exactly as expected for a star with a large CM \citep{2022MNRAS.516.2812C}. However, only Full-Frame Images are available, the extraction of light curves from which is beyond the scope of this work. We therefore elected not to include this target. The present study therefore includes the best 4 of the available 6 candidates.

Photometry for HD 182180  was obtained by the \textit{K2} mission from the Mikulski Archive for Space Telescopes (\href{https://mast.stsci.edu/portal/Mashup/Clients/Mast/Portal.html}{MAST}). The NASA \textit{Kepler} \citep{2010Sci...327..977B} satellite is a $\mu$mag-precision space photometer originally intended for long duration observations to detect exoplanets via the transit method. The photometer has a 110 square degree field of view and operates in the $400 - 850\,{\rm nm}$ bandpass. The \textit{K2} \citep{Howell_2014} mission began following the failure of \textit{Kepler's} reaction wheels. Incoming solar wind allowed \textit{Kepler} to be stabilized and observe fields along the ecliptic for roughly 3 months at a time. HD 182180 was observed by \textit{Kepler} in long-cadence mode (270 stacked 6 second exposures) during campaign 7, with observations lasting from 2015 October 04 to 2015 December 26.

$\sigma$ Ori E and HD 345439 were observed by \textit{TESS} \citep{2015JATIS...1a4003R}, a space photometer intended to search for transiting exo-planets, using four cameras with a $24^\circ\times96^\circ$ field of view and 21`` pixel scale, covering the $600-1050\,{\rm nm}$ wavelengths. \textit{TESS}'s two year nominal mission began in 2018, in which 13 sectors were covered each year, observed for 27 days each. \textit{TESS}'s mission was extended and has since re-observed a number of the same sectors.

$\sigma$ Ori E was observed by TESS in sector 6 from 2018 December 15 to 2020 December 16 with a two-minute cadence. $\sigma$ Ori E \textit{TESS} data were acquired via MAST. $\sigma$ Ori E \textit{TESS} data is contaminated and diluted by light from the nearby O star $\sigma$ Ori AB. To account for this, it is necessary to scale the data. It is difficult to infer \textit{a priori} the extent to which $\sigma$ Ori AB contaminates the \textit{TESS} light curve of $\sigma$ Ori E. The light curve of $\sigma$ Ori E has been presented in previous literature \citep{1977ApJ...216L..31H,1978ApJ...224L...5L,2010ApJ...714L.318T,2015MNRAS.451.2015O}, which all show a $\sim15\%$ difference between the minimum and maximum flux. In order to achieve the same difference in flux we compare these published data to the \textit{TESS} data and approximate that light from $\sigma$ Ori AB makes up $\sim$80$\%$ of the flux in the $\sigma$ Ori E light curve. Therefore we subtract 0.8 from the normalized \textit{TESS} light curve and then re-normalize using the mean residual flux.


HD 345439 was observed by \textit{TESS} in sector 41 from 2021 July 24 to 2021 August 20 during \textit{TESS's} extended mission. \textit{TESS}'s two minute cadence was used. \textit{TESS} photometric data were obtained via MAST.



Photometric data for HD 142184 was taken by the Microvariability and Oscillations of STars \citep[\textit{MOST};][]{2003PASP..115.1023W} satellite. These data were previously analyzed by \cite{2012MNRAS.419.1610G}. The Canadian Space Agency \textit{MOST} satellite was a space photometer dedicated to asteroseismology, in which photometry was obtained using a visible-light dual-CCD camera and a 15-cm aperture Maksutov telescope. Photometric measurements of HD 142184 were done by aperture photometry on a $20\times20$ pixel subgrid on the \textit{MOST} CCD photometer. Observations were done in switched-target mode, with part of the orbital phase being shared with observations of Arcturus. Each observation of HD 142184 contained 157 stacked 0.18 second exposures.


The Python package \textsc{Lightkurve} \citep{2018ascl.soft12013L} was used to process the photometric data of each star, with outlier rejection, binning, and normalization done on each light curve using \textsc{Lightkurve's} built-in algorithms. The rotational periods for $\sigma$ Ori E  \cite[$1.1908228_{-0.0000010}^{+0.0000012}$ d;][]{2010ApJ...714L.318T}, HD 142184 \cite[$0.508276_{-0.000012}^{+0.000015}$ d;][]{2012MNRAS.419.1610G}, HD 182180 \cite[$0.521428\pm0.000006$ d;][]{2008Rivinius} and HD 345439 \cite[$0.7701\pm0.0003$ d;][]{Wisniewski_2015} were adopted from the literature without modification, as these periods phase the data adequately well. While $\sigma$ Ori E's rotational period is known to be gradually increasing \citep{2010ApJ...714L.318T}, we elected to ignore this effect as we are not attempting to achieve a coherent phasing of the {\em TESS} data with other datasets. 

\section{Model Fitting Techniques} \label{model}

Let us now discuss fitting models from \cite{Berry_2022} to the light curves of the stars discussed in Section \ref{obs}. The model light curves themselves are grounded in the same physics as presented by \cite{Berry_2022}. Photometric modulation in these models is due to both absorption and electron scattering emission as a result of a co-rotating CM. The magnetic field in these models is a centered, tilted dipole, with contributions from higher order magnetic multipoles neglected. The light curves can be modified by changing four main parameters: magnetic obliquity $\beta$, viewer inclination from the rotation axis $i$, optical depth at the Kepler radius $\tau_{\rm K}$, and critical rotation fraction $W \equiv V_{\rm rot}/V_{\rm orb}$ where $V_{\rm rot}$ is the star's equatorial rotational velocity and $V_{\rm orb}$ is the star's surface orbital velocity.

As can be seen in the models presented by \cite{Berry_2022}, variation of $\beta$, $i$, $\tau_{\rm K}$ and $W$ have different effects on light curve morphology. Varying $\beta$ affects peak structure (rounded vs.\ flat), as well as eclipse width. Generally, low $i$ results in little photometric variation. Double eclipses arise when $i\geq60^\circ$. Further increasing $i$ deepens eclipses. $\tau_{\rm K}$ affects both absorption and emission, with larger $\tau_{\rm K}$ giving both greater levels of absorption and emission. $W$ controls the amount of both absorption and emission, eclipse depth and width, as well as peak structure. Increasing $W$ leads to light curves with greater emission, but less absorption, along with wider eclipses and more rounded peaks. 

Other parameters were used to control the structure of the magnetosphere, including a \say{latitudinal scale length} $\chi$ \citep[discussed further by][]{Berry_2022}, which controls the azimuthal distribution of density in the CM. The ratio between thermal energy and gravitational escape energy at $R_{\rm K}$, denoted $\epsilon$, is also utilized. This effectively controls the disk thickness and scale height $h_{\rm m} \approx \sqrt{2\epsilon/3}$. For all models shown in the present work, we fixed $\chi = 0.05$ and $\epsilon = 0.01$ \citep[see][for further details]{Berry_2022} in accordance with 3D MHD simulations of CMs presented by \cite{2021mobs.confE..33U}. The chosen value of $\epsilon$ gives an associated scale height $h_{\rm m} = 0.082\,R_{\rm K}$.

The light curve models presented by \cite{Berry_2022} spanned the ranges of $30^\circ \leq \beta \leq 75^\circ$, $30^\circ \leq i \leq 90^\circ$, $0.5 \leq \tau_{\rm K} \leq 2$ and $0.25 \leq W \leq 0.75$. In the present work it was deemed necessary to expand the grid of these parameters, as certain stars had best fit models that were on the edge of the grid. As such, we expanded the grid to $10^\circ \leq \beta \leq 88^\circ$, $0.1 \leq \tau_{\rm K} \leq 4$, $0.1 \leq W \leq 0.75$. New values of viewer inclination $i$ smaller than $30^\circ$ were found to be unnecessary, in accordance with the expectation that each light curve will have relatively large inclination angles.

In order to perform a precise fitting of the light curves, the grid needed to be sampled with small step sizes. However, calculating models from first principles is rather slow and expensive due to having to solve the radiative transfer equation for electron scattering \citep[see][]{Berry_2022}. Therefore we implemented an interpolation algorithm which allowed us to quickly generate a model of any combination of the four parameters that falls within the parameter grid. To do this we used the spline interpolation from the \textit{Series} functions from the \textsc{Pandas} Python package. 


To perform the fitting with our models and the observations, the light curves for each star are first normalized by \textsc{Lightkurve} to the median flux value. However, the RRM-CBO models we use for the fitting are normalized to the base flux of the star if there were no magnetosphere present. Since there is no guarantee that the median flux of the measured light curve reflects the unobstructed light of the star, it was necessary to shift the empirical light curves vertically in flux to ensure the best fit possible with each model. Given that a number of these stars exhibit a primary and secondary eclipse, the data is shifted such that the minimum values of both the data and the models are equal, so that the primary eclipse is prioritized to be fit over any secondary eclipses. Furthermore, we opted to have phase 0 coincide with the minimum flux in the star's light curves. In our models phase 0 is set to occur when the magnetic field axis is pointing towards the observer, which does not coincide with a photometric minimum. Because of this, the models are shifted in phase to ensure the best fit between the models and the data. After these shifts are performed, we use reduced $\chi^2$ (henceforth $\chi^2_{\rm red}$) in order to determine goodness-of-fit between our models and observations. Errors are found directly from the 1$\sigma$ contours in our $\chi^2_{\rm red}$ landscapes presented in Section \ref{Results}.

\begin{table*}
\caption{Best-fit parameters for each star from the current work, as well as known parameters from \protect\cite{Shultz_2019}. The \emph{$\tau_{\rm K},$ best-fit}  column refers to $\tau_{\rm K}$ as a \textit{free parameter}, whereas the \emph{$\tau_{\rm K}{\rm,\, eq.\,\ref{tk}}$} column refers to $\tau_{\rm K}$ calculated from equation \ref{tk} using parameters from \protect\cite{Shultz_2019}. For the case of $\sigma$ Ori E,  \protect\cite{Shultz_2019} assumed an inclination $i \sim 80^{\circ}$ in accordance with \protect\cite{2015MNRAS.451.2015O}. The ``\protect\cite{Shultz_2019} revised'' row makes no assumptions about $i$ and allows it to be a free parameter. Errors in the \emph{Current Work} rows are derived directly from the 1$\sigma$ contours in the $\chi^2_{\rm red}$ maps of Figures \ref{fig:37479_lc}, \ref{fig:HD142184_fit}, \ref{fig:182180_lc} and \ref{fig:HD345439_fit}  } \label{table:values}
\begin{tabular}{|p{3.2cm}|p{1.2cm}|p{1.2cm}|p{1.2cm}|p{1.2cm}|p{1.2cm}|p{1.3cm}|p{1.5cm}|p{1.5cm}|}
\hline
\hline
 &  $M_\ast (M_\odot)$& $R_\ast (R_\odot)$& $B_\ast{\rm (kG)}$&$\beta$ $(^\circ)$ & $i$ $(^\circ)$ & $W$  & $\tau_{\rm K}$, best-fit & $\tau_{\rm K}$, eq. \ref{tk}\\
 \hline
 \multicolumn{9}{c}{{\bf $\sigma$ Ori E}}\\
 \hline
 Current Work & $-$ & $-$ & $-$ & $70_{-1}^{+10}$ & $63_{-1}^{+5}$ & $0.21_{-0.1}^{+0.29}$ & $3.6_{-0.3}^{+0.4}$ & $-$\\[.2cm]
 
 \cite{Shultz_2019} & $7.9_{-0.3}^{+0.2}$ & $3.39_{-0.06}^{+0.04}$ & $10_{-1}^{+2}$ & $38\pm9$ & $77\pm4$ & $0.224_{-0.004}^{+0.003}$  & $1.5_{-0.3}^{+0.6}$ & $0.20_{-0.04}^{+0.08}$ \\[.2cm]
 
 \cite{Shultz_2019}, revised & $8.3_{-0.7}^{+0.6}$ & $3.9\pm0.6$ & $9.6\pm1.2$ & $75_{-10}^{+2}$ & $52\pm7$ & $0.229_{-0.003}^{+0.100}$  & $3.90_{-1.5}^{+4.80}$ & $0.25_{-0.10}^{+0.30}$\\
 \hline
 \multicolumn{9}{c}{{\bf HD 142184}}\\
 \hline
 Current Work & $-$ & $-$ & $-$ & $30_{-4}^{+30}$ & $58_{-14}^{+2}$ &  $0.10_{-0}^{+0.41}$ & $4.0_{-1.1}^{+0}$ & $-$\\[.2cm]
 
 \cite{Shultz_2019} & $5.7\pm0.1$ & $2.8\pm0.1$ & $9\pm2$& $9\pm3$ & $64\pm6$ & $0.524_{-0.035}^{+0.064}$ & $1.00_{-0.48}^{+0.55}$ & $1.49_{-0.72}^{+0.82}$\\
 \hline
 \multicolumn{9}{c}{{\bf HD 182180}}\\
 \hline
 Current Work & $-$ & $-$ & $-$ & $84_{-30}^{+4}$ & $66_{-6}^{+19}$ & $0.40_{-0.20}^{+0.25}$ & $0.5_{-0.1}^{+0.6}$ & $-$\\[.2cm]
 
 \cite{Shultz_2019} & $6.5\pm0.2$ & $3.2\pm0.1$ & $9.5\pm0.6$ & $82\pm4$ & $53_{-5}^{+9}$ & $0.562_{-0.061}^{+0.068}$ & $0.80_{-0.25}^{+0.28}$ & $2.30_{-0.74}^{+0.81}$\\
 \hline
 \multicolumn{9}{c}{{\bf HD 345439}}\\
 \hline
 Current Work & $-$ & $-$ & $-$ & $25_{-10}^{+9}$ & $67\pm2$ & $0.20_{-0.10}^{+0.05}$ & $2.2_{-0.6}^{+1.0}$ & $-$\\[.2cm]
 \cite{Shultz_2019} & $8.3\pm0.8$ & $3.7\pm0.5$ & $8.9\pm1.1$ & $46\pm13$ & $59\pm10$ & $0.407_{-0.076}^{+0.093}$ & $3.2_{-2.0}^{+2.3}$ & $0.89_{-0.55}^{+0.63}$ \\
 \hline
 \hline

\end{tabular}
\end{table*}

Following our model fitting, we compare our best fit parameter values to those determined directly from spectropolarimetric and spectroscopic measurements and evolutionary modelling presented by \cite{2018MNRAS.475.5144S,2019MNRAS.485.1508S,Shultz_2019}. Since our models do not directly provide the surface magnetic field strength, this comparison involves calculating the associated $\tau_{\rm K}$ for electron scattering. The appendices of \cite{2005MNRAS.357..251T} provide a CBO scaling for the peak surface density of the CM at $R_{\rm K}$:

\begin{equation} \label{sig_K}
    \sigma_{\rm K} = c_{\rm f}\frac{B^2_{\rm K}}{4\pi g_{\rm K}}\,,
\end{equation}

\noindent where $B_{\rm K} = B_{\rm eq}(R_\ast/R_{\rm K})^3$ and $g_{\rm K} = GM_\ast/R_{\rm K}^2$ are respectively the equatorial magnetic field strength and gravitational acceleration at $R_{\rm K}$ for a star with mass $M_\ast$. $c_{\rm f}$ is a correction factor found via comparison with MHD simulations, by which \cite{2020MNRAS.499.5366O} found $c_{\rm f}\approx0.3$. Furthermore, \cite{Berry_2022} introduced a surface density scaling for the magnetosphere:

\begin{equation} \label{surface_density}
    \sigma(r,\theta_{\rm o}) = \sigma_{\rm K}\biggl(\frac{r}{R_{\rm K}}\biggr)^{-p}\exp{(-\cos^2{\theta_{\rm o}}/\chi)}
\end{equation}

\noindent where $\theta_{\rm o}$ is the co-latitude, $\sigma_{\rm K}$ is defined by Equation \ref{sig_K}, and $\chi$ is the latitudinal scale length introduced above. $p$ is the radial drop-off index, where $p=5$ in the models presented by \cite{Berry_2022} and used here. This, along with the canonical electron scattering opacity $\kappa_{\rm e} = 0.344\,{\rm cm^2/g}$, and the fact that the polar magnetic field strength is twice that of the equatorial field strength (i.e. $B_* = 2B_{\rm eq,\ast}$), leads us to an expression for $\tau_{\rm K}$:

\begin{equation} \label{tk}
\tau_{\rm K} = \kappa_{\rm e}\sigma_{\rm K} = c_{\rm f}\kappa_{\rm e}\frac{B_\ast^2}{16\pi g_\ast}\biggl(\frac{R_{\rm K}}{R_\ast}\biggr)^{-4} = c_{\rm f}\kappa_{\rm e} \frac{B_\ast^2}{16\pi g_\ast} W^{8/3}\,,
\end{equation}

\noindent where $g_\ast = GM_\ast/R_\ast^2$ is the surface gravitational acceleration for a star with mass $M_\ast$ and radius $R_\ast$. The right hand side of Equation \ref{tk} uses the relation \citep{2008MNRAS.385...97U}:

\begin{equation}
    R_{\rm K} = W^{-2/3} R_\ast\,.
\end{equation}

Equation \ref{tk} shows that $\tau_{\rm K}$ is a function of $M_\ast$, $R_\ast$, $W$, and $B_\ast$, each of which we can find in the tables of Appendix C by \cite{Shultz_2019} and shown again here in Table \ref{table:values}. For a correction factor $c_{\rm f} = 0.3$, Equation \ref{tk} shows that CMs around rapidly rotating magnetic B stars should typically be marginally optically thick (i.e. $\tau_{\rm K}$ is of order unity) in the continuum. Furthermore, we see that $\tau_{\rm K}$ is dependent on the square of the surface polar magnetic field strength, which in the case of a simple dipole declines as $B_* \sim r^{-3}$. Thus the optical depth in the continuum falls off rapidly with distance (i.e. $\tau_{\rm K} \sim B_\ast^2 \sim r^{-6}$) and so most magnetospheric emission remains close to $R_{\rm K}$.

\section{Light Curve Fitting} \label{Results}

In this section we present our best fit light curves for $\sigma$ Ori E, HD 142184, HD 182180 and HD 345439, as well as $\chi^2_{\rm red}$ contour plots to visualize degeneracies in parameter space and the associated fit uncertainties. Along with our best-fit models, we show for comparison forward modeled light curves calculated using the parameters presented by \cite{Shultz_2019}, and provided here in Table \ref{table:values}. Note that $\tau_{\rm K}$ is a free parameter in the forward modeled light curves, rather than $\tau_{\rm K}$ found from Equation \ref{tk} which gives $\tau_{\rm K}$ values that are either much too small or much too large for a reasonable fit. 

All models under-fit the data, with $\sigma$ Ori E having the best-fit overall with $\chi^2_{\rm red} \approx 10$. Furthermore, the range of $\chi^2_{\rm red}$ is quite large for most stars, with $1 \lesssim \log(\chi^2_{\rm red}) \lesssim 6$.  The best-fit parameters found here and the associated uncertainties are reflective of the limitations of the simple tilted dipole model, and should not be interpreted as indicative of the actual geometrical and rotational properties of the target stars.

The contour plots presented later in this section have been re-normalized by the absolute minimum $\chi_{\rm red}^2$ ($\chi_{\rm red,min}^2$). Therefore these contour plots show $\log(\chi_{\rm red}^2/\chi_{\rm red, min}^2)$. The 1, 2 and 3$\sigma$ contours in each panel show $\log(\chi_{\rm red, min}^2+1,\,4,\,9)$, respectively. Given that re-normalization is applied first, the 1, 2 and 3$\sigma$ contours essentially show levels $\log(2)$, $\log(5)$ and $\log(10)$. Errors are derived from the 1$\sigma$ contours. Combinations of parameters that fall within the 1$\sigma$ contours should be viewed as fits of similar quality to the reported best-fit curve.



\subsection{$\sigma$ Ori E}

$\sigma$ Ori E is a very young \citep[age less than 1 Myr;][]{2022A&A...657A..60S} rapidly rotating magnetic B2Vp helium-rich star, and the prototypical example of a star whose photometric variation is mainly due to the presence of a rigidly rotating circumstellar magnetosphere \citep{2005ApJ...630L..81T}. This star hosts a relatively strong ($\sim 10\,{\rm kG}$) oblique \citep[$47^\circ \leq \beta \leq 59^\circ$;][]{2015MNRAS.451.2015O} magnetic field and a relatively fast \citep[$W \approx 0.2$;][]{Shultz_2019} rotation speed. Viewer inclination is expected to be high given the large range of variation seen with $\sigma$ Ori E, indeed quite close to $90^\circ$ \citep{2015MNRAS.451.2015O}. With this, the continuum optical depth of the CM around $\sigma$ Ori E should be large enough for electron scattering to provide a small but significant amount of emission in its light curve. Indeed, the light curve of $\sigma$ Ori E shows what may be a small emission bump at phase 0.6 \citep[see][]{1977ApJ...216L..31H,1978ApJ...224L...5L,2005ApJ...630L..81T}, recently reproduced by \cite{2022A&A...659A..37K}, and the main point of motivation for the work done by \cite{Berry_2022}.

\begin{figure*}
    \centering
    \includegraphics[width = \linewidth]{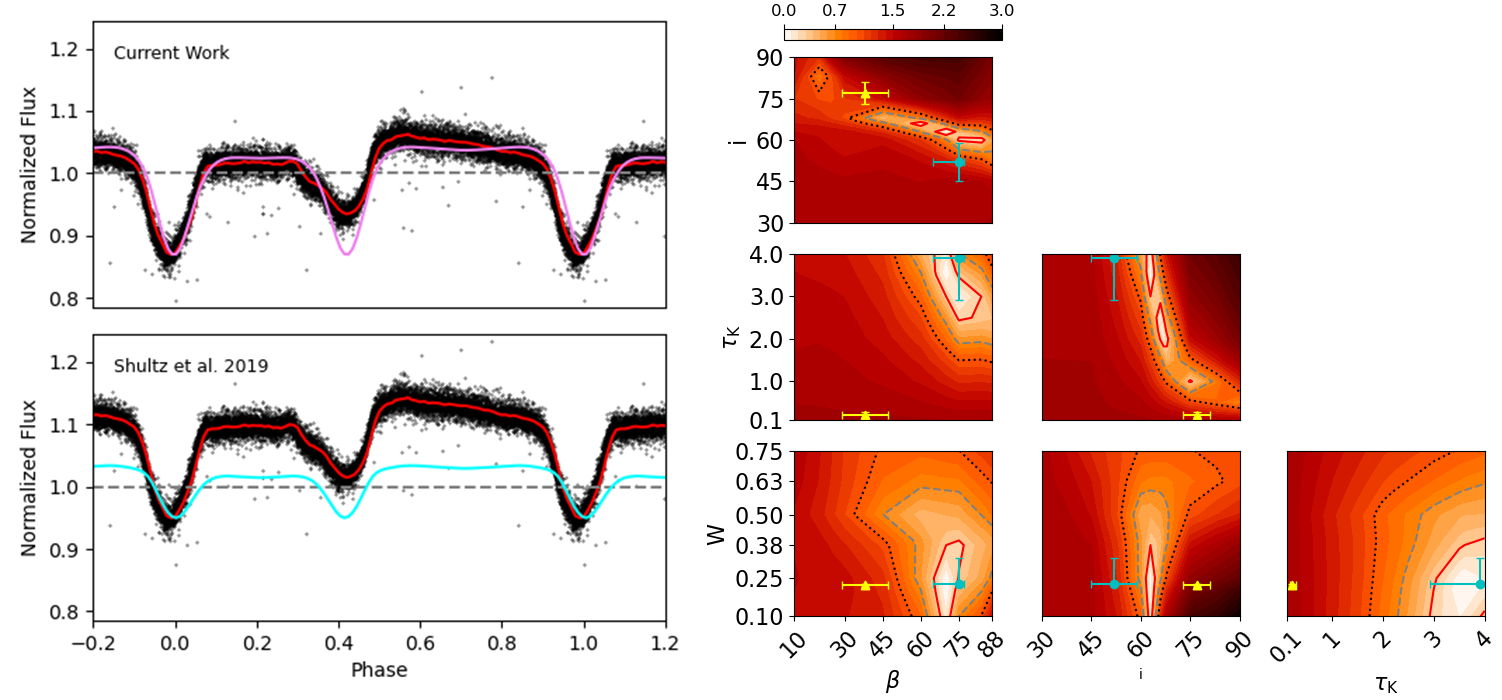}
    \caption{\textit{Left panels}: \textit{TESS} light curve of $\sigma$ Ori E (black points) with the 100 point binned light curve (red) overlaid with our best fit model (purple, top panel) and the forward modeled light curve from minimally constrained parameters \protect\citep[cyan, bottom panel;][]{Shultz_2019}. The best fit model has parameters $\beta = 70^\circ$, $i = 63^\circ$, $\tau_{\rm K} = 3.6$ and $W = 0.21$. The forward modeled curve has parameters $\beta = 75^\circ$, $i = 52^\circ$, $\tau_{\rm K} = 3.9$ and $W = 0.229$. The dashed grey line in both panels is the continuum flux, i.e. the flux from the star if there were no magnetosphere present. \textit{Right}: Contour plots showing normalized $\log$ $\chi^2_{\rm red}$ in each plane within the parameter space. Brighter areas are associated with a $\chi^2_{\rm red}$ closer to the minimum $\chi^2_{\rm red}$, and a better fit. $\log$ $\sigma$ confidence levels are shown as contour lines with $1\sigma$ (solid red), $2\sigma$ (dashed grey) and $3\sigma$ (dotted black). Values and errors directly from \protect\cite{Shultz_2019} are shown as the yellow triangles, and values found from a minimally constrained calculation are shown as the cyan points. $\tau_{\rm K}$ is calculated from Equation \ref{tk} for the yellow triangles, but is a free parameter for the cyan points. }
    \label{fig:37479_lc}
\end{figure*}

Figure \ref{fig:37479_lc} shows the \textit{TESS} light curve of $\sigma$ Ori E, along with the rotational phase-binned light curve, and the best-fit model along with the associated continuum line - i.e. the base flux of the star if no magnetosphere were present. The best fit model has parameters $\beta = 70^\circ$, $i = 63^\circ$, $\tau_{\rm K} = 3.6$, and $W = 0.21$. The primary eclipse is fit relatively well, with absorption and emission in both the model and \textit{TESS} data occurring at the same phases. The large $\beta$ associated with our model fit exhibits \say{plateaued} peaks of emission, rather than rounded peaks. This complements well with what we see in $\sigma$ Ori E, where phases between eclipses are characterized by flat-topped maxima. Indeed, these plateaus in the light curve of $\sigma$ Ori E are not of equal length and height. This effect is produced by our model light curve, and supports the idea that the obliquity of $\sigma$ Ori E's magnetic field may be higher than originally anticipated.

Another unexpected result of this analysis is the location of the continuum line with respect to the light curve of $\sigma$ Ori E. Here, both phases between the eclipses are above the continuum line, which shows that a significant amount of electron scattering emission may be occurring between eclipses. In previous analyses the continuum was assumed to be located at the level of the secondary maxima, which occurs at a flux level $\sim 2\%$ above the continuum. This then leads to a small amount of emission at phase $\sim 0.6$, which was originally picked out and the main point of motivation for the work done by \cite{Berry_2022}. With our best fit model we now show that more continuum emission may be present in the light curve of $\sigma$ Ori E than first expected.

Note that the secondary eclipse of $\sigma$ Ori E is not well fit by our best-fit model. This is a consequence of the purely dipolar magnetic field topology adopted by the RRM-CBO model used here, which leads to a plasma distribution that is symmetrical about the magnetic field axis. Hence both eclipses will always have equal depth in our models. This is unlike $\sigma$ Ori E, which is known to have a magnetic field with considerable contributions from higher-order multipoles \protect{\citep[][]{2015MNRAS.451.2015O,2018MNRAS.475.5144S}}. These contributions lead to asymmetries in the structure of the magnetosphere, which in turn leads to light curve features such as unequal eclipse depth.

A key difference between our best-fit model here and results from previous analyses is the inclination angle. The aRRM analyis presented by \cite{2015MNRAS.451.2015O} found that an inclination angle $i = 85^\circ$ is required to reproduce the eclipse spacing. Instead, we produce here a best-fit light curve with $i = 63^\circ$. Furthermore, the $\chi^2_{\rm red}$ map of Figure \ref{fig:37479_lc} shows that inclination is fairly well constrained around $i = 63^\circ$ within the $1\sigma$ confidence level.

$W$ is a parameter that is well recovered for the case of $\sigma$ Ori E. Our best fit value of $W = 0.21$ is in good agreement with \cite{Shultz_2019}, who provide $W \approx 0.22$.

The $\tau_{\rm K} = 3.6$ found here is quite substantial for a continuum optical depth, but not surprising given the amount of absorption and emission required to match the levels seen in this particular light curve. Indeed, \cite{Berry_2022} predicted that a best-fit light curve for $\sigma$ Ori E would require $\tau_{\rm K}$ to be at least 2. With $\sigma$ Ori E's $\sim 10$ kG magnetic field and relatively small $W$, Equation \ref{tk} gives $\tau_{\rm K} = 0.2$, about a factor of 10 lower than the optical depth we need to reproduce the degree of photometric variation seen in this light curve.


Due to the large discrepancies in angular values, we also compare to a set of \say{revised} values calculated using the same method and priors as adopted by \cite{Shultz_2019}, only now lifting the constraint on inclination which \citeauthor{Shultz_2019} forced to be as close to $85^\circ$ as possible in accordance with \cite{2015MNRAS.451.2015O}. As can be seen in Figure \ref{fig:37479_lc}, leaving inclination as a free parameter results in much better agreement in $i$ and $\beta$. (see also Table \ref{table:values}).

The forward modeled light curve in Figure \ref{fig:37479_lc} has parameters from the \cite{Shultz_2019} revised row of Table \ref{table:values}. Despite the close agreement in parameters, the empirical light curve is not well-fit by the forward modeled light curve. This is due to the moderate inclination $i = 52^\circ$ recovered from evolutionary modelling, which cannot produce eclipses of the required depth. This illustrates that even relatively small changes in $i$ (in this case, a mere 10$^\circ$) can lead to substantial differences in light curve amplitude. 

\subsection{HD 142184}

HD 142184 is a magnetic B2V star \citep{2012MNRAS.419.1610G} with a dipolar magnetic field strength of around $9$ kG \citep{Shultz_2019}. It has the most rapid rotation of any known magnetic early B-type star to date, with a period of $0.508275_{-0.000012}^{+0.000015}$ days \citep{2012MNRAS.419.1610G}. With a strong magnetic field and rapid rotation, this star is expected to host a large CM. The strong H$\alpha$ emission supports this \citep{2012MNRAS.419.1610G}. \cite{2012MNRAS.419.1610G} inferred a viewer inclination $i\sim70 \pm 10^\circ$. The magnetic field is marginally oblique with $\beta \sim 10^\circ$ \citep{2012MNRAS.419.1610G,Shultz_2019}. Other effects such as gravity darkening and oblateness from rapid rotation could have small effects on photometric variation from this star \citep{2012MNRAS.419.1610G}. Via visual comparison of the {\em MOST} light curve to the photometric RRM-CBO models developed by \cite{2008MNRAS.389..559T} and to the H$\alpha$ equivalent width curve, \cite{2012MNRAS.419.1610G} argued that the magnetosphere is the main cause of this star's light curve morphology.

\begin{figure*}
    \centering
    \includegraphics[width = \linewidth]{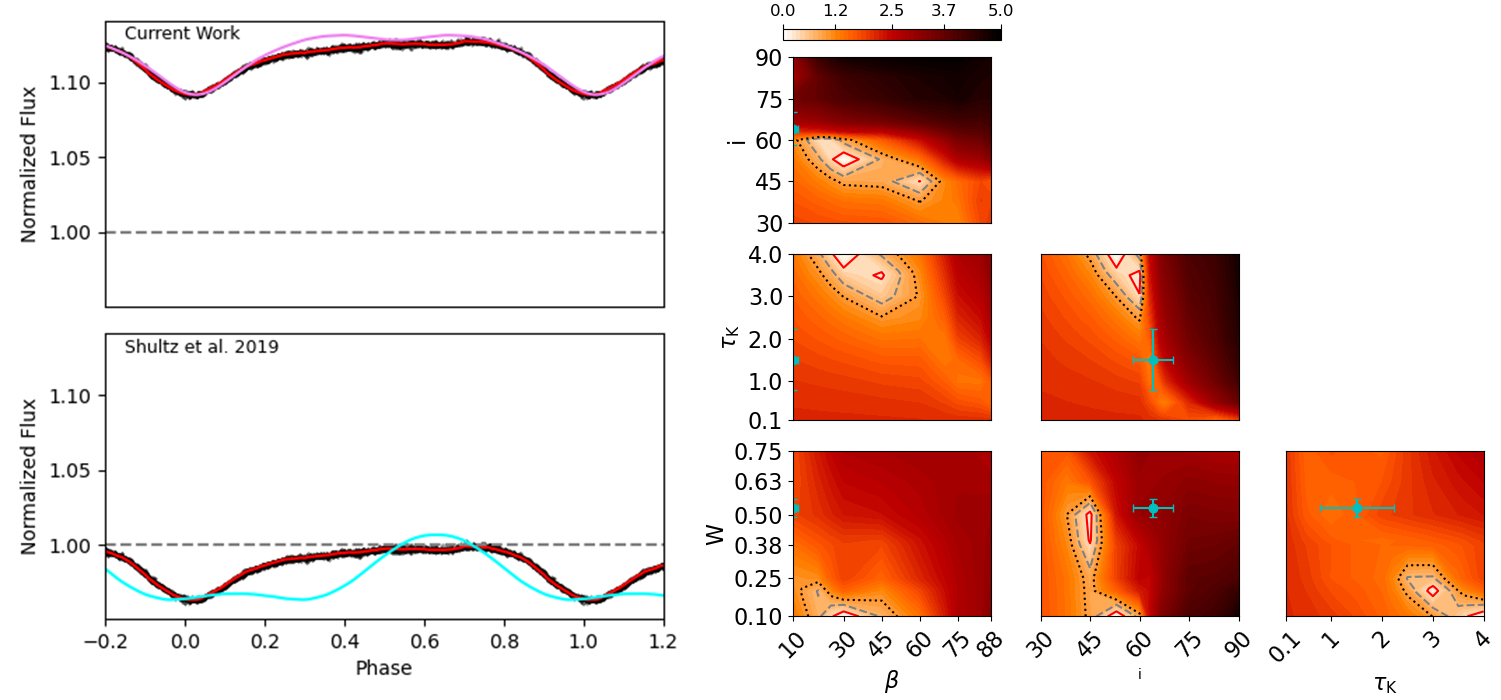}
    \caption{\textit{Left panels}: \emph{MOST} light curve of HD 142184 (black points) with the 100 point binned light curve (red) overlaid with our best fit model (purple, top panel) and the forward modeled light curve from minimally constrained parameters \protect\citep[cyan, bottom panel;][]{Shultz_2019}. The best fit model has parameters $\beta = 30^\circ$, $i = 58^\circ$, $\tau_{\rm K} = 4$ and $W = 0.1$. The forward modeled curve has the same parameters provided by \protect\cite{Shultz_2019} with a best-fit $\tau_{\rm K} = 1$. The dashed grey line in both panels is the continuum flux, i.e. the flux from the star if there were no magnetosphere present. \emph{Right}: As Figure \ref{fig:37479_lc}, except the cyan points are now taken directly from \protect\cite{Shultz_2019}. $\tau_{\rm K}$ is calculated from Equation \ref{tk}.}
    \label{fig:HD142184_fit}
\end{figure*}

Figure \ref{fig:HD142184_fit} shows the \textit{MOST} light curve (black points) with the binned light curve fitted with our best-fit model (purple). The model has parameters $\beta = 30^\circ$, $i = 58^\circ$, $\tau_{\rm K} = 4$ and $W = 0.1$. To start, our best-fit model finds $W$ considerably lower than the value determined from the star's rotation period, mass, and radius, which is about 0.5. The light curve of HD 142184 contains only one apparent eclipse. This is consistent with the H$\alpha$ emission which \cite{2012MNRAS.419.1610G} noted is asymmetrical and displays only one eclipse. \citeauthor{2012MNRAS.419.1610G} further observed that the light and H$\alpha$ equivalent width curves followed one another very closely. Our oblique models always contain two main clumps of material, which leads to double eclipsing. However, with certain combinations of our four parameters, it is possible for the stellar disk to be occulted only once by the magnetosphere. One substantial eclipse occurs at low obliquity, moderate inclination, low $W$ and higher $\tau_{\rm K}$ \citep[figure 8 of][]{Berry_2022}, which we see here. The presence of only a single eclipse in the light curve is therefore the probable reason for the large discrepancy in $W$. 

\begin{figure}
   \includegraphics[width = \columnwidth]{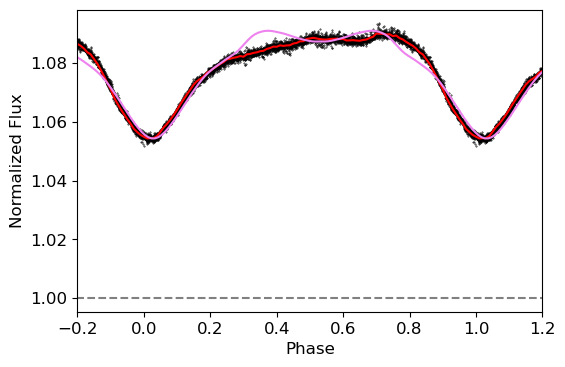}
    \caption{{\textit{MOST} light curve of HD 142184 (black points) with the 100 point binned light curve (red) overlaid with a model with parameters $\beta = 30^\circ$, $i = 45^\circ$, $\tau_{\rm K} = 4$ and $W = 0.5$ (violet). These parameter values fall within the 1$\sigma$ contour in the bottom middle panel of the corner plot in Figure \ref{fig:HD142184_fit}. This model demonstrates significantly improved agreement in terms of $W$ with \protect\cite{Shultz_2019} at the expense of fit quality and agreement in $i$. }}
    \label{W50}
\end{figure}

Fits with $W \sim 0.5$ exist within the 1$\sigma$ contour assuming $i \approx 45^\circ$ and using the same $\beta$ and $\tau_{\rm K}$ values as the best-fit curve shown in Figure \ref{fig:HD142184_fit}. Figure \ref{W50} presents an alternative fit with $\beta = 30^\circ$, $i = 45^\circ$, $\tau_{\rm K} = 4$, and $W = 0.5$, applied to the HD142184 \textit{TESS} light curve. While this alternative model shows a significant improvement in agreement with \cite{Shultz_2019} in terms of $W$, it deviates from the best-fit model regarding $i$ and offers a slightly inferior fit overall. Specifically, this alternative model also exhibits a singular apparent eclipse and reproduces the eclipse depth and width similarly to the best-fit model but does not closely match the plateaued peak width. 



\begin{figure*}
    \centering
    \includegraphics[width = \linewidth]{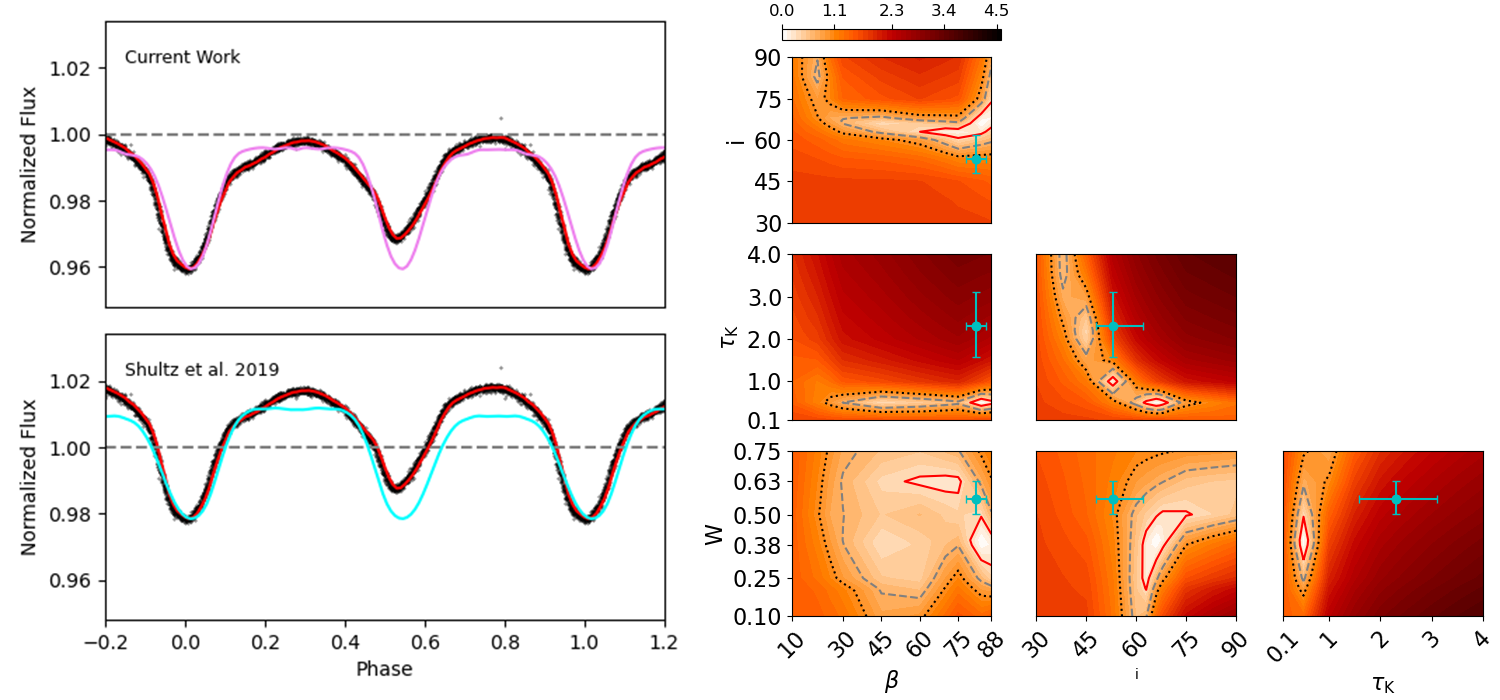}
    \caption{\textit{Left panels}: \textit{K2} light curve of HD 182180 (black points) with the 100 point binned light curve (red) overlaid with our best fit model (purple, top panel) and the forward modeled light curve from minimally constrained parameters \protect\citep[cyan, bottom panel;][]{Shultz_2019}. The best fit model has parameters $\beta = 84^\circ$, $i = 66^\circ$, $\tau_{\rm K} = 0.5$ and $W = 0.4$. The forward modeled curve has the same parameters provided by \protect\cite{Shultz_2019}, with a best-fit $\tau_{\rm K} = 0.8$. The dashed grey line in both panels is the continuum flux, i.e. the flux from the star if there were no magnetosphere present. \textit{Right}: As Figure \ref{fig:HD142184_fit}.}
    \label{fig:182180_lc}
\end{figure*}

We recover the low obliquity with our best-fit model, though it is somewhat larger than $\beta \approx 10^\circ$ obliquity determined by \cite{Shultz_2019}. The $\chi^2_{\rm red}$ map of Figure \ref{fig:HD142184_fit} shows that $\beta$ is mainly constrained around $\beta = 30^\circ$, though some higher obliquity models exist within the 1$\sigma$ contours.

We recover the moderate inclination angle determined by \cite{2012MNRAS.419.1610G} and \cite{Shultz_2019}, despite it being a bit shallower than expected. $i$ is highly constrained on the $\chi^2_{\rm red}$ map of Figure \ref{fig:HD142184_fit} around $i = 53^\circ$

$\tau_{\rm K} = 4$ for our best-fit model is sizeable for a continuum optical depth, even for the most rapidly rotating star known. However, the CM could be less optically thick with a lower limit of $\tau_{\rm K} = 2.9$ possible within the $1\sigma$ contour.



Another interesting result from HD 142184 is the location of the continuum flux with respect to the light curves. Our best-fit model shows that the flux is always $\sim 10\%$ above the continuum flux due to magnetospheric emission. As demonstrated by \cite{2012MNRAS.419.1610G}, strong H$\alpha$ emission is present at all rotational phases, which is consistent with the continuous presence of photometric emission.


Despite having parameters much closer to their inferred value, the forward modeled light curve for HD 142184 does not give a good fit. Because of the low obliquity, moderate inclination, and high $W$, the magnetosphere is usually eclipsing the star, which then leads to wide eclipses and a single rounded peak. This is the opposite of what we see in the data, where one short eclipse is present, followed by a large plateau. 

\subsection{HD 182180}

HD 182180 is a B2V helium-rich star known to host a strong \citep[$9.5\pm0.6\,{\rm kG}$;][]{Shultz_2019} oblique dipolar magnetic field \citep[$\beta = 75\pm10^\circ$, $\beta = 82\pm4^\circ$;][respectively]{2013Rivinius, Shultz_2019}, and is one of the fastest rotators known with a period of $0.521428\pm 0.000006$ days \citep{2010Oksala,2008Rivinius,2010Rivinius,2013Rivinius}. Due to the presence of a strong magnetic field and fast rotation, HD 182180 should most likely host a centrifugal magnetosphere, and this is supported by the presence of strong H$\alpha$ emission \citep{2013Rivinius}. Furthermore, the combination of a strong magnetic field and fast rotation gives a continuum optical depth of order unity, which should provide a small but significant amount of electron scattering emission. The expected large inclination, combined with a high magnetic obliquity and a relatively optically thick magnetosphere means that the amount of photometric variation in the light curve of HD 182180 may be due to magnetospheric absorption and scattering. HD 182180 is known to have sources of variability other than the magnetosphere, such as surface abundance spots \citep{2010Oksala, 2013Rivinius}. Other effects such as gravity darkening and oblateness from rapid rotation may have small effects on the photometric variation for this particular star \citep{2013Rivinius}.

The left panel of Figure \ref{fig:182180_lc} shows the {\it K2} light curve of HD 182180 over-plotted with our best-fit model (purple) with parameters $\beta = 84^\circ, i = 66^\circ, \tau_{\rm K} = 0.5$ and $W = 0.4$. The fit matches the depth of the primary eclipse well, and matches both minima and maxima in the light curve in phase. The model overestimates the depth of the secondary eclipse. This is a consequence of the tilted dipolar magnetic field topology adopted in our models - both eclipses will have the same depth as the plasma distribution in the magnetosphere is symmetrical across the magnetic field axis. Another difference between the data and our models are the rounded peaks of HD 182180's light curve, contrasted with the \say{plateaus} in our model. These plateaued peaks are an effect seen with large $\beta$ \citep[see Figures 7 and 8 of][]{Berry_2022}, as well as with slower rotation speeds, as the magnetosphere will spend more time off the limb of the stellar disk. In any case, the obliquity of HD 182180's magnetic field and its rotation speed are both known to be quite high, which is in good agreement with our values for $\beta$ and $W$ \citep{2008Rivinius,2010Oksala,Shultz_2019}.

The viewer inclination $i = 66^\circ$ obtained from our best fit model is in reasonable agreement with values derived elsewhere \citep{2008Rivinius,2010Oksala,Shultz_2019}. 


Another discrepancy between the data and our best-fit model is that the primary eclipse width is not exactly matched. This is likely due to our best-fit model having a smaller $W$ than expected, which results in sharper eclipses. Indeed, the forward modeled light curve, which has a higher $W = 0.56$, does fit the primary eclipse width better than our best-fit model.

$\tau_{\rm K}$, however, is a parameter that remains difficult to predict. Using values for the magnetic field and rotation speed provided by \cite{Shultz_2019}, ($B_\ast = 9.5$ kG, $W = 0.56$) as well as the stellar mass and radius gives an associated $\tau_{\rm K} \approx 2.3$. However, our best fit model has $\tau_{\rm K} = 0.5$, which undercuts this calculated value by a factor of $\sim 5$, and is rather unexpected given the strong magnetic field and fast rotation speed associated with this star. In fact, this is the opposite problem that we have seen for $\sigma$ Ori E, in which the predicted $\tau_{\rm K}$ is higher than the calculated value. Here, the predicted $\tau_{\rm K}$ is lower than what we calculate from Equation \ref{tk}. We are unlikely to find a good fit higher or lower than $\tau_{\rm K} = 0.5$. The $\chi^2_{\rm red}$ map of Figure \ref{fig:182180_lc} shows that $\tau_{\rm K}$ is highly constrained to $\sim$0.5 within the 1$\sigma$ confidence level.

\begin{figure*}
    \centering
    \includegraphics[width = \linewidth]{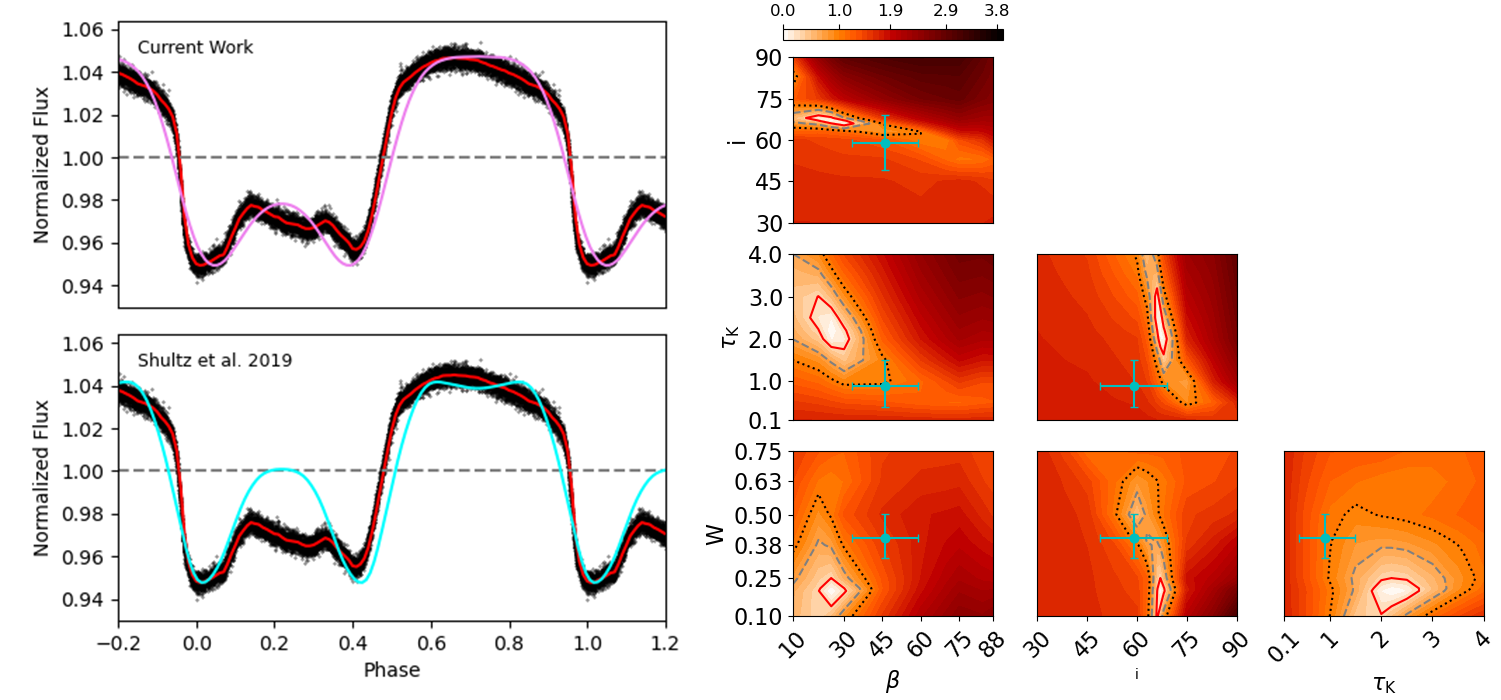}
    \caption{\textit{Left panels}: \textit{TESS} light curve of HD 345439 (black points) with the 100 point binned light curve (red) overlaid with our best fit model (purple, top panel) and the forward modeled light curve from minimally constrained parameters \protect\citep[cyan, bottom panel;][]{Shultz_2019}. The best fit model has parameters $\beta = 25^\circ$, $i = 67^\circ$, $\tau_{\rm K} = 2.2$ and $W = 0.2$. The forward modeled curve has the same parameters provided by \protect\cite{Shultz_2019}, with a best-fit $\tau_{\rm K} = 3.2$. The dashed grey line in both panels is the continuum flux, i.e. the flux from the star if there were no magnetosphere present. \textit{Right}: As Figure \ref{fig:HD142184_fit}.}
    \label{fig:HD345439_fit}
\end{figure*}

Another surprising result from our best-fit model is the location of the continuum line with respect to the observations. When looking at the light curve for HD 182180, it's quite clear to see that magnetospheric emission could be present, mainly due to the rounded peaks of this particular light curve. Photometric emission is also expected from the strong H$\alpha$ emission. However, our best fit model shows that no amount of emission above the continuum flux is present.

The forward modeled light curve in cyan is a reasonable fit, which is unsurprising given the similarity between our best-fit parameters and the values provided by \cite{Shultz_2019}. The main difference between the forward modeled light curve and our best-fit is the presence of emission in the former. This is likely due to a smaller $i$, larger $\tau_{\rm K}$ and larger $W$. While the forward modeled light curve does not match the width or depth of the secondary eclipse, this is almost certainly a short-coming of the tilted dipole model. The H$\alpha$ emission is also notably asymmetrical, probably reflecting contributions from higher-order multipoles. 

\subsection{HD 345439}

HD 345439 is a rapidly rotating B2V star known to host a strong \citep[$\sim$ 12 kG;][]{Hubrig_2015,Hubrig_2017} magnetic field with a rapid rotational period of $\sim 0.77$ days, an intermediate case between the extremely short rotation periods of HD 142184 and HD 182180, and the longer period of $\sigma$ Ori E \citep{Wisniewski_2015}. Due to the presence of a strong magnetic field and rapid rotation, this star is capable of hosting a CM. Spectroscopy of this star presented by \cite{Wisniewski_2015} shows that the equivalent widths of multiple emission and absorption lines are phase dependent. Specifically, periodic behavior of the star's double-peaked H$\alpha$ emission indicates that a CM exists around HD 345439, as originally inferred by \cite{2014ApJ...784L..30E} on the basis of the similar morphology of the star's Brackett lines. \cite{Wisniewski_2015} suggest that photometric variability of HD 345439 is due in most part to a co-rotating magnetosphere, but do not rule out the possibility that pulsations have an effect on the light curve morphology. \cite{2020MNRAS.499.5379S} noted that HD\,345439 has the strongest H$\alpha$ emission in the sample of CM host stars, further supporting the possibility of a magnetospherically dominated light curve. The {\em TESS} light curve was examined by \cite{2022ApJ...924L..10J}, who suggested that its variability was likely entirely consistent with the presence of a CM.

\cite{Wisniewski_2015} suggested that this star should have a high inclination ($i\sim 75^\circ$) and a moderate magnetic obliquity ($\beta\sim 45^\circ$) based on photometric RRM light curves presented by \cite{2008MNRAS.389..559T}. Spectropolarimetric and evolutionary modelling by \cite{Shultz_2019} found $i=59 \pm 10^\circ$, $\beta = 46 \pm 13^\circ$, and $B_{\rm d} = 9 \pm 1$~kG, consistent with the predictions of \cite{Wisniewski_2015} and \cite{Hubrig_2017}.

Figure \ref{fig:HD345439_fit} shows the \textit{TESS} light curve along with our best-fit model in purple, with parameters $\beta = 25^\circ$, $i = 67^\circ$, $\tau_{\rm K} = 2.2$ and $W = 0.2$. The fit matches the overall morphology of this light curve quite well, with the levels of emission and absorption occurring at the same level and time in phase. The light curve exhibits two maxima of highly unequal fluxes, with one below the grey dashed continuum line, and one above. This effect in the model is most likely due to the fact that the magnetosphere is still occulting part of the stellar disk when the magnetic field axis is pointing away from the observer, an effect from the relatively low $\beta$ and $W$. 

The primary maximum of the light curve of HD 345439 displays a \say{plateaued} peak, similar to those seen in the light curve of $\sigma$ Ori E. This feature is somewhat matched by our model, though the duration of the plateau does not last as long in phase as it does in the \textit{TESS} data. The secondary maximum of the observations also displays a plateaued peak. This is not matched by the secondary maximum of our best-fit model, which exhibits a rounded peak. The length of these plateaus can be increased by increasing $\beta$ or by decreasing $W$. $W$ is already quite small, so it's unlikely that it should be decreased. A larger $\beta$ may be in order in accordance with $\beta \approx 45^\circ$ as \cite{Wisniewski_2015} and \cite{Shultz_2019} have suggested. 

However, increasing $\beta$ will also have the effect of increasing the flux of the secondary maximum, which would most likely lead to an overestimation of the secondary maximum seen for HD 345439. Furthermore, the $\chi^2_{\rm red}$ map of Figure \ref{fig:HD345439_fit} shows that $\beta$ is indeed  constrained around $\beta = 25^\circ$. $W$ is also constrained around our best fit value of $W = 0.2$ with a range of $0.1 \lesssim W \leq 0.25$ within the $1\sigma$ confidence level. This shows that it is certainly possible for $W$ to be somewhat larger, but not large enough to be in agreement with the value of $W = 0.4 \pm 0.1$ found by \cite{Shultz_2019}.

Inclination $i = 67^\circ$ agrees well with values found by \cite{Wisniewski_2015} and \cite{Shultz_2019}. The $\chi^2_{\rm red}$ map of Figure \ref{fig:HD345439_fit} shows that inclination for this star is tightly constrained around $i = 67^\circ$.

The optical depth $\tau_{\rm K} = 2.2$ for our best-fit model is sensible, given the amount of absorption and emission required to reasonably fit the light curve of HD 345439. The associated $\tau_{\rm K} = 0.9$ calculated from Equation \ref{tk} using values  found by \cite{Shultz_2019}. The $\chi^2_{\rm red}$ map of Figure \ref{fig:HD345439_fit} shows that $\tau_{\rm K}$ is fairly constrained, with a range $1.6\leq \tau_{\rm K} \leq 3.2$ within the $1\sigma$ confidence level.  

The location of the continuum flux (grey dashed line) with respect to the light curves shows that there is a significant level of magnetospheric emission occurring around HD 345439, with $\sim 4\%$  extra emission from magnetospheric scattering.

The forward modeled light curve for this star is quite similar in structure to our best-fit model, with the main differences being a result of a larger $\beta$ and $W$ than our best fit. $\beta = 46^\circ$ results in a higher and wider secondary peak. Meanwhile, the larger $W$ results in a slightly skinnier absolute maximum. $\tau_{\rm K} = 3.2$ in the forward modeled light curve is even larger than the best-fit model. However, given the smaller $i$ and larger $W$ than in the latter, a larger $\tau_{\rm K}$ in the former is needed to match the levels of absorption and emission in the \textit{TESS} light curve.

\section{Discussion} \label{Discussting}

\begin{figure}
    \centering
    \includegraphics[width = \columnwidth]{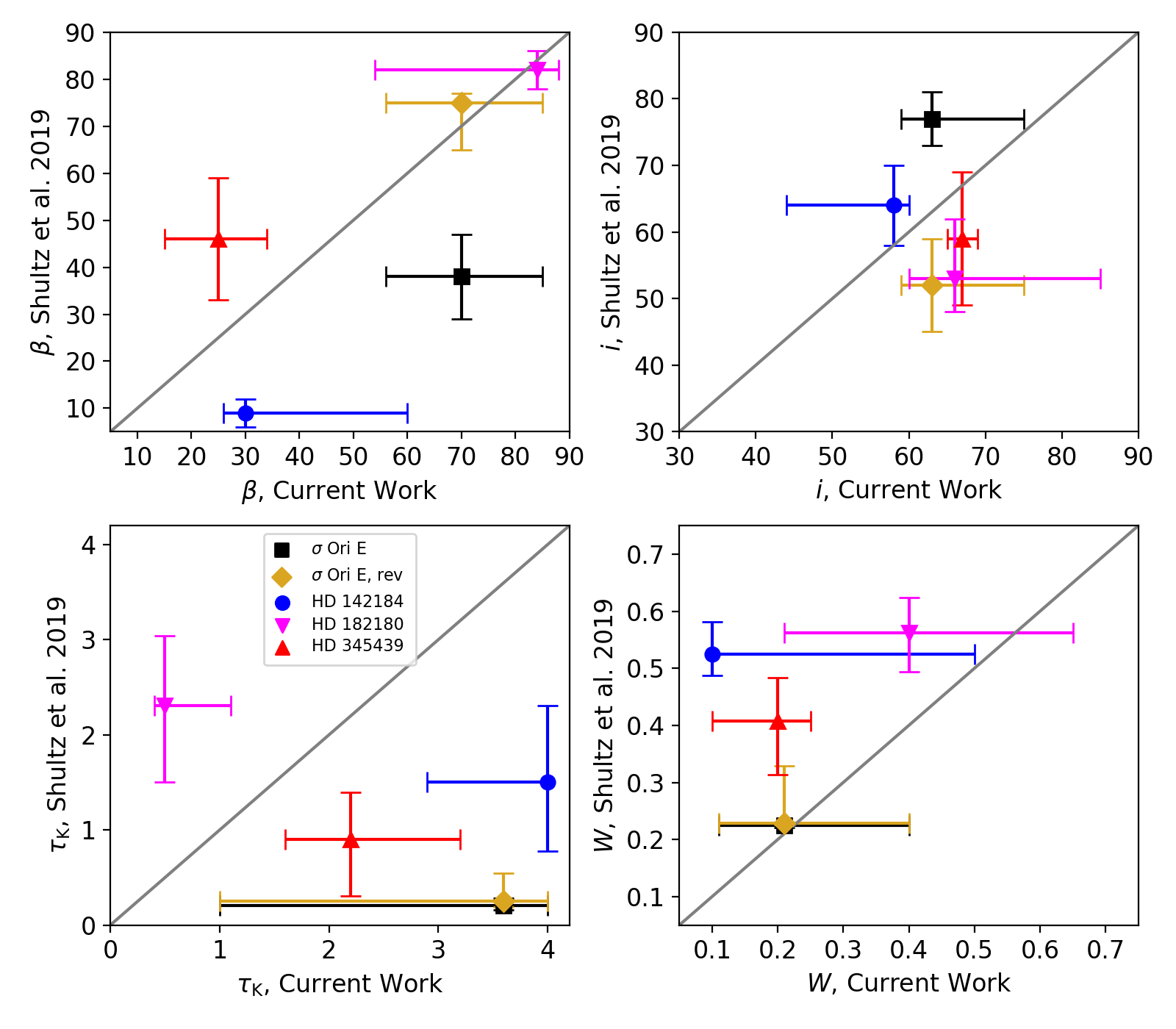}
    \caption{$\beta$, $i$, $\tau_{\rm K}$, and $W$ comparisons between our best-fit models and values found by \protect\cite{Shultz_2019}, and listed in Table \ref{table:values}. $\tau_{\rm K}$ is calculated from Equation \ref{tk}. The closer each point is to the grey line, the better the agreement.}
    \label{fig:shultz_v_berry}
\end{figure}

Our aim in this work was to do a blind test on a sample of stars, using only the available photometry, along with models presented by \cite{Berry_2022} and expanded upon here in order to infer magnetic and rotational properties of these stars. Parameters inferred from the photometric analysis were then compared to those independently determined by \cite{Shultz_2019} via spectroscopic measurements of atmospheric parameters and projected rotation velocities, direct spectropolarimetric measurements of the surface magnetic field, rotational periods determined from photometric, spectroscopic, and spectropolarimetric data, and stellar parameters inferred from evolutionary models. Figure \ref{fig:shultz_v_berry} compares our best-fit values for each parameter to those determined by \cite{Shultz_2019}. We have found that the photometric RRM models, which assume a purely dipolar magnetic field topology, can in most cases approximately recover three out of the four parameters for most stars, with the notable exception of HD 142184. 

\subsection{Parameter Comparison}


There is some agreement in $\beta$ specifically for HD 182180 and the $\sigma$ Ori E revised values; however, there is a slight tendency towards higher $\beta$ than those found by \cite{Shultz_2019}. One reason for this discrepancy is the plateaued peaks which occur in the light curves of all stars other than HD 182180. The presence of plateaued peaks, rather than rounded peaks, is a signature of large $\beta$ in the models presented by \cite{Berry_2022}. The only exception to this discrepancy is HD 345439, in which we underestimate $\beta$ as compared to the value given by \cite{Shultz_2019}. Indeed, a larger $\beta$ would result in a longer plateau for this star's absolute maximum. However, increasing $\beta$ also raises the secondary maximum as seen by the cyan curve in Figure \ref{fig:HD345439_fit}. $\beta$ also has a slight effect on eclipse width, with eclipses becoming sharper with increasing $\beta$.

Fair agreement in $i$ exists between our models and the values provided by \cite{Shultz_2019}. This is expected, as a large inclination is required to produce any noticeable photometric variation \citep{2008MNRAS.389..559T,Berry_2022}. In order to recover the double eclipsing seen in each star here (with the exception of HD 142184), a combination of $i$ and $\beta$ must exist such that $i+\beta \gtrsim 90^\circ$ \citep{2008MNRAS.389..559T}. This condition is met by our models as well as with values from \cite{Shultz_2019} for all stars other than HD 142184, which helps explain why that star only shows one clear eclipse.

$W$ is approximately recovered in the sense that stars with high or low $W$ as inferred from rotational periods and evolutionary models also have high or low $W$ in the photometric models (again with the exception of HD 142184). The values yielded by our fits are systematically lower than the values found by \cite{Shultz_2019}. Once again, the light curves for stars other than HD 182180 show prominent plateaued peaks, which, as well as being a sign of lower $\beta$, is an effect of low $W$. Furthermore, larger $W$ will lead to wider eclipses, which can last $\sim0.2$ in phase for $W \geq 0.5$ \citep{Berry_2022}. However, none of the eclipses in these stars last that long in phase, which in our models is indicative of smaller $W$.


The one parameter we cannot reliably recover whatsoever is $\tau_{\rm K}$. We use Equation \ref{tk} to predict what $\tau_{\rm K}$ should be given $M_\ast$, $R_\ast$, $B_\ast$ and $W$. We do not reliably recover $\tau_{\rm K}$ using our models. This was an issue first mentioned in the discussion by \cite{Berry_2022}, in which they noted that $\tau_{\rm K} \sim 2$ would be needed to match the levels of photometric variation seen in the light curve of $\sigma$ Ori E. However, using the same expression as Equation \ref{tk}, \cite{Berry_2022} found $\tau_{\rm K} \approx 0.21$ for this star, which would not nearly match the light curve amplitude. This discrepancy is confirmed here, as our best-fit model predicts $\tau_{\rm K} = 3.6$ for $\sigma$ Ori E. Furthermore, Equation \ref{tk} underestimates the best-fit $\tau_{\rm K}$ for HD 345439 and HD 142184 as well. Oddly enough, Equation \ref{tk} \textit{overestimates} $\tau_{\rm K}$ given by our best-fit model for HD 182180.

\begin{figure}
    \centering
    \includegraphics[width=\columnwidth]{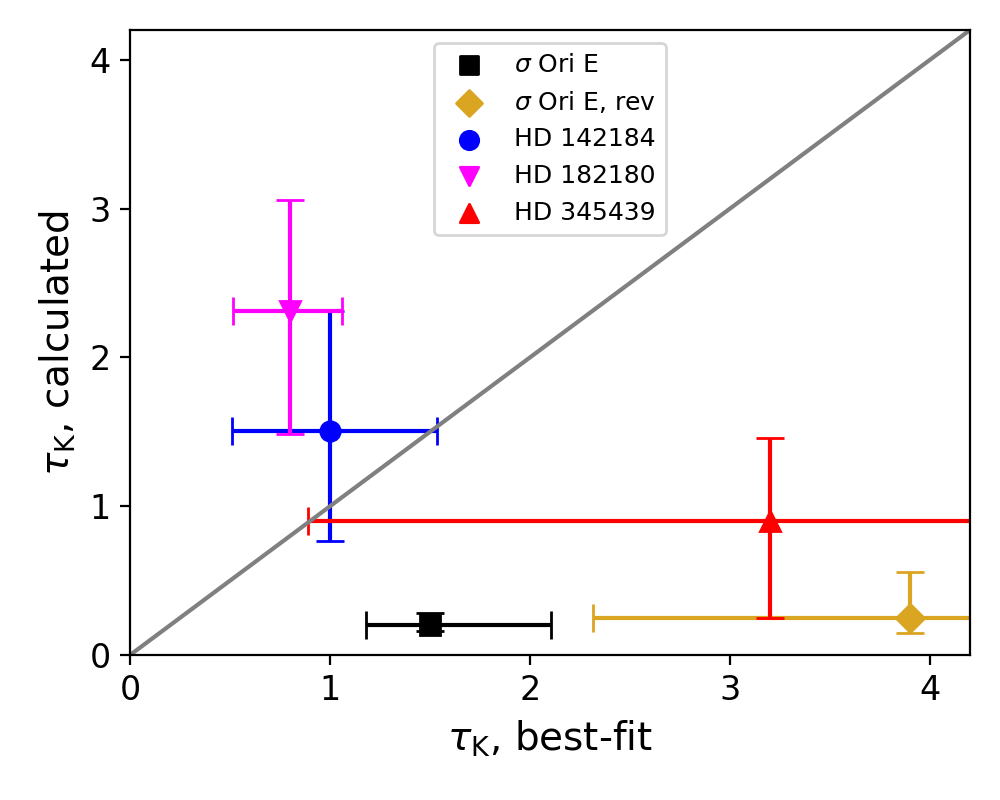}
    \caption{Calculated $\tau_{\rm K}$ from Equation \ref{tk} versus $\tau_{\rm K}$ as a free parameter using the forward modeled light curves from \protect\cite{Shultz_2019}, with the associated parameters listed in Table \ref{table:values}.}
    \label{fig:shultz_v_shultz}
\end{figure}

Some possible causes of this discrepancy were discussed by \cite{Berry_2022}. One possibility has to do with the correction factor $c_{\rm f}$, assumed to be 0.3 by \cite{2020MNRAS.499.5366O} based on calibrations from the 2D MHD simulations presented by \cite{2008MNRAS.385...97U}. This calibration may be incorrect for oblique models. For the case of $\sigma$ Ori E, \cite{Berry_2022} found that $c_{\rm f}$ needs to be of order unity. Doing the same thing for HD 345439 and HD 142184 using the best-fit $\tau_{\rm K}$ from our models and associated $M_\ast$, $R_\ast$, $B_\ast$ and $W$ from \cite{Shultz_2019}, we again find that $c_{\rm f}$ needs to be of order unity. Simply increasing $c_{\rm f}$, however, would not solve the overestimation in calculated $\tau_{\rm K}$ for HD 182180. In order to get a  calculated $\tau_{\rm K} \approx 0.5$, $c_{\rm f} \sim 10^{-2}$ is needed, an order of magnitude smaller than the value inferred by \cite{2020MNRAS.499.5366O}. The main difference between HD 182180 and the other stars in this analysis is the extremely high obliquity of $\beta \sim 80^\circ$. This leads us to propose that $c_{\rm f}$ may be a function of $\beta$ that does not monotonically increase.

Another factor which could play a role in determining $\tau_{\rm K}$ is the star's effective temperature, $T_{\rm eff}$. Figure \ref{fig:shultz_v_shultz} shows $\tau_{\rm K}$ according to Equation \ref{tk} versus the best-fit $\tau_{\rm K}$ inferred from the forward-modelled light curves. The intent behind this comparison is to control for systematic effects introduced by the differences in angular and rotational parameters between the best-fit models and the independently determined values. For $\sigma$ Ori E and HD 345439, Equation \ref{tk} underestimates the best-fit $\tau_{\rm K}$. However, for HD 142184 and HD 182180, Equation \ref{tk} {\em overestimates} the best-fit $\tau_{\rm K}$. The main difference between these two groups of stars is their $T_{\rm eff}$. Specifically, HD 142184 and HD 182180 are cool for their spectral type \citep[$T_{\rm eff}\approx 17\,{\rm kK}$ for both stars;][]{2012MNRAS.419.1610G,2013Rivinius}, while $\sigma$ Ori E and HD 345439 are hotter \citep[$T_{\rm eff} = 23\pm2\,{\rm kK}$ for both stars;][]{2019MNRAS.485.1508S}. $T_{\rm eff}$ plays a direct role in a star's mass loss rate, with cooler stars having less powerful winds. As discussed by \cite{2020MNRAS.499.5379S} and \cite{2020MNRAS.499.5366O}, a leakage mechanism competing with CBO cannot be ruled out, and such a mechanism should become more important towards lower mass-loss rates.





The ultimate discrepancy between our models and the data is the fact that our models use a purely dipolar magnetic field. This leads to symmetry in our models, in which the two clouds of material in the CM are essentially mirrored about the magnetic field axis. This in turn leads to symmetries in our model light curves (see figures 7 and 8 of \citealt{Berry_2022}, in which every light curve is mirrored about phase 0.5). Such symmetry is not seen in any of the data, with light curves having different eclipse depths and rounded or plateaued peaks. HD 142184 has only one eclipse in both the continuum and H$\alpha$, when we would expect it to have two based on its geometric properties. Based on magnetic data, both $\sigma$ Ori E and HD 142184 are known to depart from a purely dipolar geometry \citep{2015MNRAS.451.2015O,2018MNRAS.475.5144S}. Asymmetries in H$\alpha$ emission bumps seen in each star in this sample \citep{2012MNRAS.419..959O,2012MNRAS.419.1610G,2013Rivinius,Wisniewski_2015}, point to the presence of more complex magnetic field topologies than a pure dipole. 

This almost certainly explains many of the disagreements in angular and rotational parameters. For example, in the case of HD\,142184, the only way to achieve one eclipse with a purely dipolar model is for $W$ to be relatively small, such that the eclipsing clouds are located relatively far from the stellar surface. However, a non-dipolar field can introduce an asymmetry in the magnetospheric plasma distribution such that one region is much denser than the others, despite the tilt of the dipole component being small. This would then be expected to result in a single eclipse. 


Despite this, in 3/4 of the sample we can still recover $i$ and $\beta$ as well as reasonably fit these stars' light curves. This has important implications. Tilted dipoles are by far the simplest topology to model, and make the fewest assumptions. Complex magnetic field geometries can only be reliably measured using high-resolution spectropolarimetric time series, which are difficult to obtain for extremely rapidly rotating stars seen at large inclinations (which will have very broad spectral lines), especially for those which are dim \citep[e.g. with \textit{V} magnitudes $\geq 8$, one of which is HD 345439 with \textit{V} = 11.11 mag;][]{2000A&A...355L..27H}. 

\subsection{Comparison to other models}

Photometric variation from CM's governed by higher order multipoles has been investigated by \cite{2022A&A...659A..37K}, who found that the rigidly rotating magnetosphere model can explain subtle features in light curves only when higher order multipoles dominate the magnetic field out to $R_{\rm K}$. \cite{2022A&A...659A..37K} considered multipoles up to order $n = 9$, with combinations of magnetic field topologies considered as well. Similar physics to those employed by \cite{Berry_2022} were used to simulate light curves with both magnetospheric absorption and electron scattering emission. 

The light curve of $\sigma$ Ori E was fit by \cite{2022A&A...659A..37K}, using a magnetic field with a combination of a dipole and a quadrupole, and is a considerably better fit than that shown in this work. Magnetospheric absorption and emission was taken into account, as well as surface spots from \cite{2015MNRAS.451.2015O}. Both eclipses are well fit due to the introduction of quadrupole contributions, whereas we overestimate the secondary eclipse by about $5\%$. However, while we do see an emission bump in the model by \cite{2022A&A...659A..37K}, it does not match the morphology of the inferred emission bump which occurs at phase $\sim0.6$. Furthermore, \cite{2022A&A...659A..37K} adopt $\beta = 90^\circ$ for their model fit, even larger than our best-fit $\beta = 70^\circ$.

There are three key differences between the modeling approaches utilized by \cite{2022A&A...659A..37K} and that adopted here. The first is that we restricted magnetic geometries to simple dipoles. The second is that \cite{2022A&A...659A..37K} simplified radiative transfer to make it less computationally expensive to solve. Specifically, \cite{2022A&A...659A..37K} assumed that light only reflects off the surface of the clouds, with any emission inside the clouds ignored. \cite{2022A&A...659A..37K} found this approximation to be acceptable. In comparison, the RRM models developed by \cite{Berry_2022} and used in this work fully solve the radiative transfer equation using a scattering source function, which means that emission throughout the whole magnetosphere is considered. The third difference is that \cite{2022A&A...659A..37K} adopted a density scaling $\rho \sim r^{-3}$, the same scaling introduced by \cite{2005MNRAS.357..251T} which is based on the magnetosphere filling time. This leads to a slower decrease in density and overall less dense magnetosphere. The model presented by \cite{Berry_2022} and again used here instead utilized the recent CBO scalings developed by \cite{2020MNRAS.499.5366O} which have density scale like $\rho \sim r^{-5}$. This leads to a sharper decline in density, but makes the magnetosphere overall more dense than the $r^{-3}$ scaling. Our models also take recent MHD simulations \citep{2021mobs.confE..33U} into account, thus motivating the introduction of $\chi$, which controls the azimuthal distribution of density. 

The models used by \citeauthor{2022A&A...659A..37K} achieve higher quality fits as compared to those shown here, demonstrating that higher-order multipoles and spot modelling are indispensable, and have a larger impact on fit quality than radiative transfer or density scaling. The logical next step would be to combine these two models, i.e. the density scaling and radiative transfer model used in this work combined with higher-order multipoles and spot models. However, we also note that the ability to forward-model higher-order multipolar contributions to the surface magnetic geometry is dependent upon the availability of high-resolution spectropolarimetric time series data with dense coverage of the rotational phase curve. Since such data are observationally expensive, they are unlikely to be available for the majority of stars for which high-quality space photometry can be obtained. It may therefore also be of interest to see whether and to what degree multipolar RRM models can be used to infer multipole contributions to the circumstellar magnetic environment independent of spectropolarimetric constraints.



\subsection{Application to other $\sigma$ Ori E variable Stars}


Recent preliminary analysis of \textit{TESS} light curves of magnetic chemically peculiar stars by \cite{2020pase.conf...46M} shows that several objects have \say{warped} light curves, characterized by a large number of harmonics in their periodogram. These light curves could not be fully reproduced via chemical spot models. These stars have parameters predicting large CMs \citep[e.g.][]{Shultz_2019} and have H$\alpha$ emission properties consistent with this expectation \citep{2020MNRAS.499.5379S}.

One such star is HD 37776. The main reason why this star could not be analyzed here is the fact that this star is known to host a magnetic field with complex multipolar topology \citep{2011ApJ...726...24K}. As such, attempting to infer this star's magnetic and rotational properties using our dipolar models would be inappropriate. Furthermore, most of the photometric variation is due to surface spots \citep{2007A&A...470.1089K}. This star was analyzed by \cite{2022A&A...659A..37K}, who used surface spots as well as an absorbing and emitting magnetosphere dominated by an octupolar magnetic field. \cite{2022A&A...659A..37K} found that the presence of this magnetosphere accounted for subtle features in this light curve's morphology. While the light curves analyzed here could largely be reproduced with purely magnetospheric models, it's likely that the majority of stars with CMs -- which tend to have weaker H$\alpha$ emission and, presumably, less circumstellar material -- will require simultaneous modelling of surface chemical abundances and magnetospheres in order to reproduce their light curves.

Another star which could be analyzed in the future is HD 37017. This star is not expected to produce eclipses due to its fairly shallow inclination $i = 38^\circ$ and magnetic obliquity $\beta = 56^\circ$ \citep[][; see also Figure\ \ref{fig:targets}]{Shultz_2019}. Indeed, there is no sign of eclipses in H$\alpha$ \citep{2020MNRAS.499.5379S}. However, H$\alpha$ also shows that only one cloud may be present in the magnetosphere \citep{2020MNRAS.499.5379S}, which is indicative of complex magnetic field topology. Alternatively, this highly asymmetric cloud structure is reminiscent of that seen in the tidally locked binary system HD\,156324, in which the gravitocentrifugal accumulation surface of the CM is warped by the orbital Lagrange geometry \citep{2018MNRAS.475..839S}. Since HD\,37017 is also a short-period binary \citep{1998A&A...337..183B}, its magnetospheric geometry might also be affected by the gravitational influence of the companion. However, in contrast to HD\,156324, HD\,37017's orbit is eccentric, and the star's rotation is not synchronized with the orbital period; if there is an orbital influence, time-independent RRM models may prove inappropriate.


The CM star Tr 16-26 very recently discovered by \cite{2022MNRAS.516.2812C} is an excellent candidate for the same analysis performed in this work. This star shows double-humped H$\alpha$ emission, hosts a strong $\sim 11$kG surface magnetic field, has a rapid rotation period of approximately 0.97 days, and portrays two apparent eclipses in its \textit{TESS} light curve which coincide with magnetic nulls. In fact, the overall morphology of this star's \textit{TESS} light curve is strikingly similar to that of $\sigma$ Ori E. However, due to its discovery during the writing of this paper and the lack of readily available space photometry, we have decided not to include Tr16-26 here. Tr16-26 should absolutely be analyzed in future work.

In general, stars suspected or confirmed of having large $\beta \sim 90^\circ$ should be focused on in future studies. HD 182180 is the only star in the current work which is known to have high magnetic obliquity, a condition that we recovered. Since this star was the only example of Equation \ref{tk} overestimating the best-fit $\tau_{\rm K}$, other stars with large $\beta$ should be analyzed to see if this phenomena is a trend for these stars, or an anomaly in the case of HD 182180.

\section{Summary and Future Work} \label{Summary}

We've used RRM models with CBO density scaling from \cite{Berry_2022}, and applied them to a sample of rapidly rotating, strongly magnetic early B-type stars known to have strong magnetospheric H$\alpha$ emission: $\sigma$ Ori E, HD 182180, HD 345439 and HD 142184. This is the largest comparison of photometric RRM models to light curves conducted to date, and includes 4 of the 5 sample stars for which high-quality space photometry is readily available. 

Previous photometric RRM studies have relied upon forward modelling of the light curve based upon known magnetic and rotational parameters, almost exclusively for $\sigma$ Ori E. By contrast, we have performed a blind test of the model grid, seeking to determine whether the best-fit models can recover the magnetic,geometrical, and rotational properties of the target stars as inferred from spectropolarimetric, spectroscopic, and evolutionary modelling.

The primary limitation of our model is its assumption of a purely dipolar magnetic field, which is known to be incorrect in two cases, and is likely to be incorrect in the others. Despite this, we find that we can approximately infer geometrical and rotational properties of $\sigma$ Ori E, HD 182180 and HD 345439. We cannot reliably infer the properties of HD 142184, as our best fit rotation parameter is far too low. This discrepancy is likely due to HD 142184's magnetosphere being strongly affected by higher-order multipole contributions. This star should be prioritized for aRRM modelling in order to test this hypothesis. 

The optical depth at the Kepler radius $\tau_{\rm K}$ is not recovered for any of the sample stars. Moreoever, the direction of the discrepancy is different for different objects: two require much higher values of $\tau_{\rm K}$ than those expectec from theory, two much lower. This may indicate that the correction factor may be a function of $\beta$ and/or $T_{\rm eff}$. 

We did not consider the contribution of surface chemical spots to the light curve. Future work should explore their impact using maps inferred from Doppler Imaging of the most important chemical species. In the majority of cases chemical spots are expected to be the dominant source of photometric variation, however their influence is not likely to be entirely zero even in the cases of the stars examined here. 

We fixed the azimuthal density scaling factor $\chi$ to the MHD simulation-calibrated value adopted by \cite{Berry_2022}. Future work should explore the results of varying this parameter. Broadband polarimetry may provide a powerful independent constraint on $\chi$. As demonstrated by \cite{2013ApJ...766L...9C}, the scattering geometry inferred from polarimetric measurements of $\sigma$ Ori E could not be reconciled with a simultaneous fit to the photometric time series using an RRM model with a filling-time density scaling. Revisiting the simultaneous constraints provided by photometric and polarimetric time series using the CBO-modified RRM model may prove fruitful. Such modelling was done for the Of?p star HD\,191612 by \cite{2022MNRAS.511.3228M}, with excellent results. 

RRM models can also be used to predict H$\alpha$ emission properties. Simultaneous modelling of photometry with H$\alpha$ could provide additional important constraints on the magnetospheric plasma distribution. 

Despite our use of the simplest possible magnetospheric model, we can still approximately recover $\beta$, $i$ and $W$ for 3 of the 4 stars. This indicates that tilted dipole models may be a means by which to roughly infer the geometrical and rotational properties of stars with CMs whose apparent magnitudes place them beyond the reach of contemporary spectropolarimetry. The severe limitation on fit quality introduced by the tilted dipole assumption may well be a source of systematic error in the inference of other quantities ($i$ and $W$ in particular). Since the high-resolution spectropolarimetric time series data required to infer the surface magnetic geometry via tomographic inversion are not likely to be available for most stars, it may be of interest to use RRM models with multipole expansions to determine to what degree multipole geometries can be reliably inferred on the basis of photometry alone, comparing these to those stars such as $\sigma$ Ori E for which spectropolarimetrically constrained magnetic models are available.



\section*{Acknowledgements}

The authors thank Jason Grunhut for providing the {\it MOST} data for HD 142184. The authors thank their colleagues in the MOBSTER collaboration for the thoughtful feedback provided during internal review. MES acknowledges the financial support provided by the Annie Jump Cannon Fellowship, supported by the University of Delaware and endowed by the Mount Cuba Astronomical Observatory. AuD acknowledges support by NASA through Chandra Award number TM1-22001B
and GO2-23003X issued by the Chandra X-ray Observatory 27 Center, which
is operated by the Smithsonian Astrophysical Observatory for and on
behalf of NASA under contract NAS8-03060. SPO and AuD acknowledge support from NASA ATP grant 80NSSC22K0628.

\section*{Data Availability Statement}
The software used is available upon request from the authors. {\em TESS} and {\em K2} data are available on the MAST archive. {\em MOST} data are available upon request from the authors.

\bibliographystyle{mnras}
\bibliography{export-bibtex.bib}

\end{document}